\documentstyle[emulateapj]{article}
\submitted{To appear in AJ}

\lefthead{\sc Blakeslee}
\righthead{\sc Dense Cluster Globulars}

\makeatletter
\newenvironment{inlinefigure}{%
\def\@captype{figure}%
\noindent\begin{minipage}{0.999\linewidth}\begin{center}}
{\end{center}\end{minipage}\smallskip}
\makeatother

\newcommand\kms{km~s$^{-1}$}
\newcommand\kmsMpc{km~s$^{-1}$~Mpc$^{-1}$}
\newcommand\hmpc{$h^{-1}\,$Mpc}
\newcommand\hkpc{$h^{-1}\,$kpc}
\newcommand\etal{{et~al.}} 
\newcommand\doph{{\sc DoPhot}} 
\newcommand\sn{\ifmmode S_N\else$S_N$\fi}
\newcommand\siglf{\ifmmode \sigma_{\sc lf}\else$\sigma_{\sc lf}$\fi}
\newcommand\sigcl{\ifmmode \sigma_{\rm cl}\else$\sigma_{\rm cl}$\fi}
\newcommand\egc{\ifmmode{\eta_{\sc{gc}}}\else$\eta_{\sc{gc}}$\fi}
\newcommand\pf{\ifmmode{{\hbox{\sc psf}}}\else{{\sc psf}}\fi}
\newcommand\erfc{{\hbox{erfc}}}
\newcommand\mbar{\ifmmode\overline m\else$\overline m$\fi}
\newcommand\mM{\ifmmode(m{-}M)\else$(m{-}M)$\fi}
\newcommand\mo{\ifmmode m^0\else$m^0$\fi}
\newcommand\Mo{\ifmmode M^0\else$M^0$\fi}
\newcommand\mor{\ifmmode m^0_R\else$m^0_R$\fi}
\newcommand\mstar{\ifmmode{m^*_1}\else{$m^*_1$}\fi}
\newcommand\mfifty{\hbox{$m$\rlap{\lower 1pt\hbox{$_R$}}$^{50\%}$}}
\newcommand\pgc{\ifmmode P_{\rm GC}\else$P_{\rm GC}$\fi}
\newcommand\lta{\mathrel{\rlap{\lower 3pt\hbox{$\mathchar"218$}}
     \raise 2.0pt\hbox{$\mathchar"13C$}}}
\newcommand\gta{\mathrel{\rlap{\lower 3pt\hbox{$\mathchar"218$}}
     \raise 2.0pt\hbox{$\mathchar"13E$}}}
\newcommand\msun{\ifmmode{\hbox{M$_\odot$}}\else{M$_\odot$}\fi}

\begin{document}

\title{Globular Clusters in Dense Clusters of Galaxies\altaffilmark{1}}

\author{John P. Blakeslee}
\affil{Palomar Observatory, California Institute of Technology,
Mail Stop 105-24, Pasadena,~CA~91125\\ 
Electronic mail: {\tt jpb@astro.caltech.edu}}
\authoremail{jpb@astro.caltech.edu}
 
\altaffiltext{1}{Based on observations obtained at the 
W.M. Keck Observatory, operated as a scientific partnership
by the California Institute of Technology, the University of California,
and the National Aeronautics and Space Administration.
Keck Observatory was made possible by the generous
financial support of the W.M.~Keck Foundation.
}

\begin{abstract}

Deep imaging data from the Keck~II telescope are employed to study the
globular cluster (GC) populations in the cores of six rich Abell clusters.
The sample includes A754, A1644, A2124, A2147, A2151, and A2152 and covers
the redshift range $z = 0.035$--0.066.  These clusters also span a range 
in morphology from spiral-rich, irregular systems to centrally concentrated
cD clusters rich in early-type galaxies.  Globular cluster specific
frequencies \sn\ and luminosity function dispersions \siglf\ are measured
for a total of 9 galaxies in six central fields.  The measured values of
\sn\ for the six brightest cluster galaxies (BCGs) are all higher than
typical values for giant ellipticals, in accord with the \sn-density
correlations found by Blakeslee, Tonry, \& Metzger (1997).  The three
non-BCGs analyzed also have elevated values of \sn, confirming that central
location is a primary factor.  Two different models are used to estimate
\egc, the number of GCs per unit mass, for these central cluster fields.
The values for \egc\ are consistent with those found in the earlier sample,
again indicating that the number of GCs scales with mass and that the \sn\
variations are due to a deficit of halo light, i.e., \sn\ reflects
mass-to-light ratio.  The similarity of the GC color distributions of BCGs
and more ordinary ellipticals further implies that no special mechanism is
needed for explaining the properties of these GC populations.

The discussion builds on an earlier suggestion that the GCs (both metal
rich and metal poor) around the central cluster galaxies were assembled
at early times, and that star formation halted prematurely in the
central galaxies at the epoch of cluster collapse.  It is argued that
this ``BCG saturation'' model is consistent with recent simulations of
BCG/cluster formation.  The subsequent addition of luminous material to
the BCG through cluster dynamical evolution can cause \sn\ to decrease
while conserving \egc, but both theory and observations indicate that
the time scale for this is long.  However,
there may be some evidence of it in the present sample and elsewhere.  
Finally, the GC luminosity function measurements are used to constrain
the relative distances of the three clusters that
make up the Hercules supercluster.
\smallskip
\end{abstract}


\section{Introduction}

Enormous populations of 10,000 or more globular clusters (GCs) 
surround the central giant galaxies in rich galaxy clusters.
This was first noticed more than forty years ago in the
case of M87 at the center of the Virgo cluster
(Baum 1955). While a GC specific frequency \sn\
(number of GCs per unit $M_V{\,=\,}{-}15$ of galaxy luminosity;
Harris \& van den Bergh 1981) near 4 is typical for giant elliptical
galaxies, some galaxies such as M87 have $\sn\sim12$.
Harris \& Petrie (1978) concluded that either (1)~the number 
of GCs scales with total mass and M87 has a mass-to-light
ratio $M/L$ that is 2-3 times higher than other giant ellipticals,
or (2) local initial conditions at the center of the Virgo cluster
stimulated M87 to form more GCs per unit mass.

Until recently, the sparse data available on the subject
indicated that \sn\ for central cluster galaxies
was uncorrelated with any other obvious galaxy or cluster properties
(Harris, Pritchet, \& McClure 1995; West \etal\ 1995).
The puzzling variations in \sn\ among these galaxies
was interpreted as reflecting local initial conditions, and therefore
unpredictable from our current vantage point.
Thus, the latter of Harris \& Petrie's two explanations 
grew in popularity (e.g., Harris 1991; McLaughlin, Harris \& Hanes 1994).
A general trend of increasing \sn\ with environmental density 
was recognized, however, and various authors attempted to explain
why galaxies in denser environments, and
central cluster galaxies in particular, might have
been more efficient in forming GCs
(e.g., West 1993; Harris \& Pudritz 1994).

A recent study of a complete sample of central galaxies in 19 Abell
clusters within 10,000~\kms\ showed that, in contrast to the
conclusions of previous works, \sn\ for these galaxies correlates well
with overall cluster properties, including velocity dispersion,
X-ray luminosity, and the local density of bright galaxies
(Blakeslee 1997).  This study used the method of Blakeslee \& Tonry
(1995), which combines information from the counts of faint point
sources in the galaxy halos with measurements of the residual surface
brightness variance after the galaxy and detected point sources are
removed.  The method allows for better constraints on the total
number of GCs while simultaneously constraining the form of the GC
luminosity function (GCLF), which is of interest because 
of its frequent use as a standard candle distance indicator
(see reviews by Harris 1991 and Whitmore~1996).

Blakeslee, Tonry, \& Metzger (1997, hereafter BTM) showed that
the observed \sn\ correlations were consistent with a simple model in
which the number of GCs scales with mass rather than luminosity.
(This was the alternative possibility raised by Harris \& Petrie.)
The GC formation efficiency, or the rate of GC formation per
unit mass $\egc = N_{\rm GC}/M$, would therefore be universal.
BTM found roughly $\egc \sim 0.7\;$GC per $10^9\,M_{\odot}$.
The observed correlations of central galaxy \sn\ with 
cluster density are then due to this scaling of GC number
with mass in conjunction with the relative insensitivity of
the present-day galaxy luminosity to cluster richness
(e.g., Postman \& Lauer 1995).  This is the same phenomenon
that makes these galaxies rough standard candles.
Thus, rather than having anomalously large GC populations, 
these galaxies appear to be underluminous for their prominent
positions in the centers of very rich clusters.

Blakeslee (1997) speculated that tidal disruption from the collapse of
the surrounding cluster may have removed the star-forming gas and halted
further luminosity growth in the central galaxy sometime after the GCs
had formed.  Harris, Harris, \& McLaughlin (1998) suggested that
protogalactic winds driven by a rapid initial starburst were what removed the
gas and halted the luminosity growth.  Building on this
idea, McLaughlin (1999b) has compiled further evidence for a universal
GC formation efficiency.  These ideas do not conflict with evolutionary
models in which GCs and other material are later stripped from the rest
of the cluster members and added to the central galaxy (Muzzio 1987;
C\^ot\'e, Marzke, \& West 1998).

The present work uses Keck imaging data to extend the BTM study
to larger distances and higher density environments.
Because the BTM observations were conducted on a 2.4\,m telescope,
the sample was limited to clusters within $z\lta0.033$,
and this limited the variety of clusters that could be studied.
In fact, the richest in that sample was the Coma cluster, the
very first one for which data were obtained (Blakeslee \& Tonry 1995).
Thus, although BTM found strong correlations of the
central galaxy \sn\ with cluster properties,
the full range of these correlations had yet to be explored.
The sample of galaxies presented here extends twice as far 
in distance as the earlier sample.

The following section describes the selection and properties of the present
sample of galaxies.  Observations and data reductions are discussed
in Section~3, which includes the radial number density distributions of the
faint objects that cluster around the central galaxies.  Section~4 presents
the main results on the GC luminosity functions, specific frequencies, and
inferred mass formation efficiencies.  Section~5 then discusses the
relevance of these results for formation and evolution models of GC
populations and for estimates of the distances to these clusters.  
The final section provides a summary.

\section{The Galaxy Sample}

The goal of this program was to extend the BTM survey of GC populations
in 23 giant elliptical galaxies (gE's) in 19 clusters to denser and more
diverse environments.  Because of practical considerations, the selected
clusters are all in the northern spring sky and at redshifts $z<0.07$.
Table~\ref{tab:samp} summarizes the cluster X-ray, dynamical, and
morphological properties.  All X-ray information is due to Jones \&
Forman (1999).  The velocity information comes from a variety of
sources, including Zabludoff \& Zaritsky (1995) for A754, Hill \&
Oegerle (1993) as modified by Fadda \etal\ (1996) for A2124, and Barmby
\& Huchra (1998) for A2147, A2151, and A2152.  There are discordant
values in the literature for the A1644 velocity dispersion.  Struble \&
Rood (1991) quote $\sigma_{\rm cl} = 991$ \kms\ from 92 galaxies and
Zabludoff \etal\ (1993) find $939^{+87}_{-68}$ \kms\ from 76 galaxies,
but Fadda \etal\ (1996) find a gently falling velocity dispersion
profile and quote an asymptotic value of $759^{+61}_{-56}$ \kms\ from 84
galaxies.  The tabulated value of $890$ \kms\ is found from a
3$\sigma$-clipped average of 63 galaxies from the NASA/IPAC
Extragalactic Database (NED) within 1\,\hmpc\ of the central galaxy.

There are also discrepant dispersion values published for 
A2147 and A2152, the two overlapping clusters that along with
A2151 comprise the Hercules supercluster.  
Zabludoff \etal\ (1993) found $\sigma_{\rm cl} = 1081^{+170}_{-117}$ 
for A2147 and $\sigma_{\rm cl} = 1346^{+276}_{-173}$ for A2152 
from 32 and 21 galaxy velocities, respectively;
Barmby \& Huchra (1998) find the much smaller
tabulated values using 93 and 56 velocities, respectively.
These latter authors employed a method of minimizing 
a vector that consisted of the two projected spatial
dimensions and the velocity offset
to unambiguously assign galaxies to individual clusters 
(or the ``dispersed supercluster'' component) in this complex region.
This approach may tend to yield lower dispersion values but appears
to be more consistent in giving mean velocities closer to the
measured values for the respective central brightest cluster members.
For instance, previous investigations generally found very similar
redshifts for the three Hercules clusters, while Barmby \& Huchra
find that A2152 has a significantly higher mean velocity, close
to that of its central galaxy (Postman \& Lauer 1995).

\begin{deluxetable}{cccccccccl}\tablenum{1}
\small
\centering\tablewidth{0pt}
\newdimen\digitwidth
\setbox0=\hbox{\rm0}
\digitwidth=\wd0
\catcode`?=\active
\def?{\kern\digitwidth}
\tablecaption{Abell Cluster Sample\label{tab:samp}}
\tablehead{
\colhead{Cluster} &\colhead{ ~RA$_{\sc x}$~(J2000)~Dec$_{\sc x}$}& \colhead{ $z_{\sc cmb}$}& \colhead{$\,\sigma_{\rm cl}\,$} &
\colhead{$L_{\sc x}\,$} &\colhead{$\;kT_{\sc x}$} & {R} & {BM} & {RS} &{Constellation} 
} \startdata
?A754 & ?9 09 04.3 ~~$-$09 39 38 & 0.0552 & $900 \pm 70$ & $6.89 \pm 0.10$ & 9.1 & 2 &   I & cD & Hydra \\
A1644 & 12 57 19.3 ~~$-$17 22 06 & 0.0485 & $890 \pm 85$ & $2.49 \pm 0.03$ & 4.7 & 1 & II & cD & Virgo \\
A2124 & 15 45 00.4 ~~$+$36 06 56 & 0.0660 & $880 \pm 80$ & $1.16 \pm 0.05$ & 3.5\tablenotemark{e} & 1& I & cD & Corona Borealis\\
A2147 & 16 02 15.3 ~~$+$15 57 58 & 0.0354 & $820 \pm 65$ & $1.89 \pm 0.02$ & 4.4 & 1& III &  F & Hercules \\
A2151 & 16 04 37.0 ~~$+$17 43 40 & 0.0367 & $705 \pm 45$ & $0.88 \pm 0.02$ & 3.8 & 2& III &  F & Hercules \\
A2152 & 16 05 37.8 ~~$+$16 26 17 & 0.0434 & $715 \pm 75$ & $0.21 \pm 0.02$ & 1.9\tablenotemark{e} & 1& III &  F & Hercules \\
\enddata \tablenotetext{e}{Estimated from the $L_{\sc x}$--$T_{\sc x}$ relation
of Jones \& Forman (1999), which has an uncertainty of about 55\%.
\\ \vskip -20pt} \vspace{-0.3cm}
\tablecomments{Table lists for each cluster: coordinates of the peak of the 
extended X-ray emission (Jones \& Forman 1999); redshift in the cosmic
microwave background rest frame; velocity dispersion (km~s$^{-1}$);
X-ray luminosity in the 0.5--4.5~keV band (10$^{44}$ ergs/s) and X-ray
gas temperature (keV) (Jones \& Forman 1999); richness class and
Bautz-Morgan type (Abell et al.\ 1989); Rood-Sastry type (Struble 
\& Rood 1987); and the host constellation.  See text for a 
discussion of the sources of the velocity information.  }
\end{deluxetable}

The first three clusters in Table~\ref{tab:samp} are centrally concentrated
Bautz-Morgan type I/II, Rood-Sastry type cD clusters dominated by 
a single giant galaxy, while the last three are less concentrated
BM type~III, RS type~F clusters with more ordinary
brightest members.  The three Hercules clusters also have elevated
spiral fractions, with that of A2151 being about 50\%
(Tarenghi \etal\ 1980), as compared to the 20-30\% typical 
of rich clusters (Dressler 1980a,b).
Since the velocity dispersions and (except for A754 and A2152) 
X-ray temperatures of all the sample clusters
are fairly similar, the masses might also be similar.
Thus, the observed morphological differences may be
solely indicative of evolutionary state, and one might 
look for evolutionary effects in the GC populations.

Few of the central cluster galaxies in the present sample have 
been the subjects of photographic surface photometry studies that
have classified them as ``cD galaxies,'' meaning that they have
extended envelopes with luminosities in excess of the 
$r^{1/4}$-law profiles defined by the inner parts (Oemler 1976).
Moreover, the Rood-Sastry
``cD'' classification does not depend on having a central galaxy
with this sort of significantly extended cD envelope.  For instance,
the bright central galaxy in the F-type cluster A2147 is classified
as a cD galaxy (Schombert 1988).  To avoid confusion, this paper
will generally refer to the program galaxies as ``central galaxies,''
or as ``brightest cluster galaxies'' (BCGs).
As shown most notably by M87 in the Virgo cluster
(and in other clusters discussed by BTM), the central cluster galaxy
need not be the BCG; however, the BCGs are central for all of the 
clusters in the present sample.

\begin{deluxetable}{ccccccccccc}\tablenum{2}
\small
\centering\tablewidth{0pt}
\newdimen\digitwidth
\setbox0=\hbox{\rm0}
\digitwidth=\wd0
\catcode`?=\active
\def?{\kern\digitwidth}
\tablecaption{Keck LRIS Observations of Central Cluster Galaxies\label{tab:obs}}
\tablehead{
\colhead{Field} & \colhead{ ~RA~(J2000)~Dec }& \colhead{~$l$~~~~~~~~$b$} & 
\colhead{$A_{R}$} & \colhead{Exp} & \colhead{\sc psf} & \colhead{ sec$\,z$} &
\colhead{$m_1^*$} & {$\mu_{\rm sky}$} &
{?\hbox{$m$\rlap{\lower 1pt\hbox{$_R$}}$^{50\%}$}} &  {$m^0_R$}
} \startdata
?A754 & ~?9 08 32.3 ~~$-$09 37 48 & 239.20 ~~$+$24.70 & 0.17& 4600& 0.84&1.28& 37.19& 20.9& 26.4 & 28.9\\
A1644 & ~12 57 11.8 ~~$-$17 24 35 & 304.89 ~~$+$45.44 & 0.18& 6000& 0.66&1.28& 37.47& 21.0& 27.0 & 28.6\\
A2124 & ~15 44 58.8 ~~$+$36 06 35 & ?57.76 ~~$+$52.30 & 0.07& 8400& 0.57&1.10& 37.96& 20.7& 27.3 & 29.3\\
A2147 & ~16 02 17.0 ~~$+$15 58 28 & ?28.91 ~~$+$44.52 & 0.08& 3350& 0.71&1.25& 36.94& 20.3& 26.3 & 27.9\\
A2151 & ~16 04 35.8 ~~$+$17 43 18 & ?31.47 ~~$+$44.66 & 0.11& 2800& 0.64&1.10& 36.72& 20.5& 26.5 & 27.9\\
A2152 & ~16 05 29.3 ~~$+$16 26 11 & ?29.91 ~~$+$43.99 & 0.11& 2000& 0.54&1.03& 36.37& 20.7& 26.7 & 28.3\\
\enddata
\vspace{-0.4cm}
\tablecomments{Columns list: cluster name; right ascension and declination (J2000)
of the central galaxy; Galactic longitude and latitude of the central galaxy;
$R$-band Galactic extinction in this direction (Schlegel et al.\ 1998);
total LRIS exposure time in the final image (sec);
full-width at half-maximum of the point spread function (arcsec);
mean airmass; magnitude $m_1^*$ of an object that would produce
one source count per total integration time (see text);
sky brightness in mag~arcsec$^{-2}$; the 50\% completeness limit
for point source detection (mag; see text);
and the expected $R$-band GCLF turnover magnitude.}
\end{deluxetable}

\section{Observations and Reductions}

Deep $R$-band images of the sample galaxies were obtained in April 1997 
with the Low Resolution Imaging Spectrograph (LRIS) (Oke \etal\ 1995) on
the 10\,m Keck~II telescope on Mauna Kea, Hawaii.  The image scale was
0\farcs211~pix$^{-1}$ and the usable area was about $7\farcm2\times5\farcm5$.
The median seeing was about 0\farcs65.
The moon was at last quarter, but conditions were
photometric and several Landolt (1992) standard fields were observed for
calibration.  Independent photometry for four of the program fields was
acquired at Palomar Observatory and agreed to within 0.02~mag in each case.
Individual LRIS exposures were bias-subtracted, flat-fielded, and combined,
rejecting cosmic ray hits, to produce the final images.  Two amplifier
readout was used, but otherwise this proceeded as detailed by BTM.

Table~\ref{tab:obs} provides a summary of the LRIS Abell cluster
observations.  The coordinates for the $z<0.05$ BCGs
come from Postman \& Lauer (1995); the coordinates for the other
two are from NED.  The quantity \mstar\ listed in the table is
one measure of the depth of an observation and can be compared directly
to the values listed by BTM.  Formally, it corresponds to the magnitude
of an object that is bright enough to produce one detected photo\-electron
per total image integration time, corrected for Galactic extinction.
Thus, the extinction-corrected magnitude of an 
object yielding $f$ total counts in the image is simply
$m = -2.5\log(f) + \mstar$.  
The \mfifty\
quantity in the table gives a better measure
of overall image quality; it corresponds to the 
extinction-corrected 50\% completeness limit
for object detection above a 4$\,\sigma$ threshold.
The numbers were determined from the completeness experiments
discussed below.

Figures~1--6 (appended to end of paper) display the central
$\sim\,$4\farcm2 of the deep LRIS $R$-band images.  The A2124 image
shows the redshift $z{\,=\,}0.57$ gravitationally lensed arc 27\arcsec\
to the southeast of the cD center (Blakeslee \& Metzger 1999).  To study
the GC populations, we model and subtract the galaxy halo light then
subtract the large-scale residuals before performing the point-source
photometry and image power spectrum measurements.  The reduction and
analysis methods are described in detail by BTM and Blakeslee \& Tonry
(1995).  The following subsections summarize the reduction procedure,
highlighting the steps that had to be altered for the current data set.

\subsection{Point Source Photometry and Completeness}
\label{ssec:psource}

LRIS has a spatially varying point spread function (\pf).  When stellar images
near the center of the chip are in focus, the images near the chip
corners are significantly extended.  The width of the \pf\ varies by as
much as 15\% across the chip; if unaccounted for, this could produce a
photometry error of $\sim\,$0.3~mag.  BTM used a version of the
photometry program \doph\ (Schechter \etal\ 1993) that had been modified
to account for additional noise from the galaxy light that had been
modeled and subtracted.  That version made no allowance for a variable
\pf, so we instead used the version described by Metzger \& Schechter
(1998), and kindly provided by M.~Metzger, that allows the \pf\ shape
parameters to vary as a two-dimensional quadratic function of position
on the CCD.  We modified this version so that it also allowed for extra
noise from the subtracted galaxy light and used an aperture correction
that varied as an independent two-dimensional quadratic function of
position.  With the allowance made for the variable \pf, the fitted
spatial variation of the aperture correction was typically less than
0.07~mag and always less than 0.1~mag.  In the end, we disregarded all
information on objects very near the CCD corners, more than about
3\farcm5 from the center, where the stellar images were worst.

To test the accuracy of the object magnitudes, we did aperture photometry
by hand (in the same way as for the Landolt standards) for about 15 of the
brightest, relatively clean, unsaturated stars in each image.  The results
agreed at the 0.02~mag level with the \doph\ magnitudes 
for all of the fields except A2124, for which the \doph\ values were
systematically off by $+0.07{\,\pm\,}0.02$ mag.  This field has the highest
galactic latitude and so is relatively devoid of stars.  In addition,
because of the cluster's distance and the need for a longer total integration,
the individual exposures were twice as long as for the other fields with
comparably good seeing, and stars with $m_R<22.5$ were saturated.
For these reasons, there were few good stars for doing absolute
photometry in an automated way, and the empirical correction of 0.07~mag
was applied to the magnitudes in this field. 

As in BTM, a series of artificial star experiments were performed for each
field to test the completeness limits of the point source detection 
and check for systematic biases in the recovered magnitudes as a function
of magnitude. Composite stars were constructed and added in grids 
(so as to avoid artificial crowding while minimizing the number of separate
tests required) to the images and then recovered with \doph.
The grid spacing was 30~pix, 20\% larger than the \pf\ fit box.
The tests proceeded in 10 or 11 steps of 0.4~mag, starting at 
$m_{\rm add}{\,=\;}23.0$ or $m_{\rm add}{\,=\;}23.4$.
For each magnitude in each field, two separate \doph\ runs were
performed using grids offset with respect to each other by half 
a grid spacing in each direction. This allowed for better area
sampling of the completeness function. Over 350,000 stars were
added in 124 separate \doph\ runs in the course of these experiments.
The 50\% detection completeness limits shown in Table~\ref{tab:obs}
are the values determined from these tests at radii beyond 
$\sim\,$1\arcmin\ from the central galaxy.
The number of real objects found by \doph\ in these fields ranged
from about 6,000 to nearly 10,000.  

The results of these tests were used in choosing the
cutoff magnitudes $m_c$ for each field at which the point source 
detection completeness was 85--90\% and any photometric bias in
recovered magnitudes was negligible.  As the completeness generally
depends on radius, two or three different cutoff magnitudes
were used for different radial regions in each field.  
Uncertainties in the completeness fraction 
$f_c \equiv {\rm N}_{\rm found}/{\rm N}_{\rm add}$ were calculated as
$\delta f_c = [f_c\,(1{\,-\,}f_c)/{\rm N}_{\rm add}]^{1/2}$ (Bolte 1989)
and included in the uncertainty estimated for the corrected counts.
The final results for the counts due to GCs, corrected for incompleteness 
and background, are tabulated in Table~\ref{tab:data} of~\S\ref{sec:results}.

\subsection{Radial Number Density Distributions}

After rejecting extended objects and correcting for incompleteness, 
the radial surface density of point sources in each field
was fitted to a de Vaucouleurs $r^{1/4}$~law plus background model
\begin{equation}\label{eq:devauc}
 N_{\rm ps}(r_p) \,=\, 
   N(R_e)\times\exp[-7.67(r_p/R_e)^{1/4} - 1] + N_{\rm bg} \,,
\end{equation}
where $r_p$ is the projected radial distance (referenced this way to
avoid confusion with the $R$ magnitudes), $R_e$ is the effective radius
of the distribution, and $N_{\rm bg}$ is the background point-source
level integrated over the magnitude range of the counts.  The purpose of
these fits was to estimate the backgrounds (primarily due to distant
unresolved galaxies); the same approach was employed by BTM and previous
authors (e.g., Harris 1986).
The GC counts are then computed as $N_{\rm GC} = N_{\rm ps} - N_{\rm bg}$.
Figures~\ref{fig:a754rad} through \ref{fig:a2152rad} show the
radial distributions of the point sources found by \doph\ to a
completeness level of $\sim\,$90\% and the corresponding fits.

Many different radial binnings were explored for each field, and 
an average background value was chosen from among the fits.
When the innermost points significantly changed the fit (due to
a leveling off of the counts at small radii, as for A2152
in Fig.~\ref{fig:a2152rad}), they were excluded.  
Although one might consider adding other radial components to the model
(for faint unresolved cluster dwarfs or a distinct population of
intergalactic GCs), the quality of the fits ($\chi^2/N\sim1$) does
not warrant it.  Any such additional components must
be at a low level or have a spatial distribution extended enough 
to be nearly flat over the range of interest. \nl\nl

\setcounter{figure}{6}
\begin{inlinefigure}\medskip 
\centerline{\includegraphics[width=0.90\linewidth]{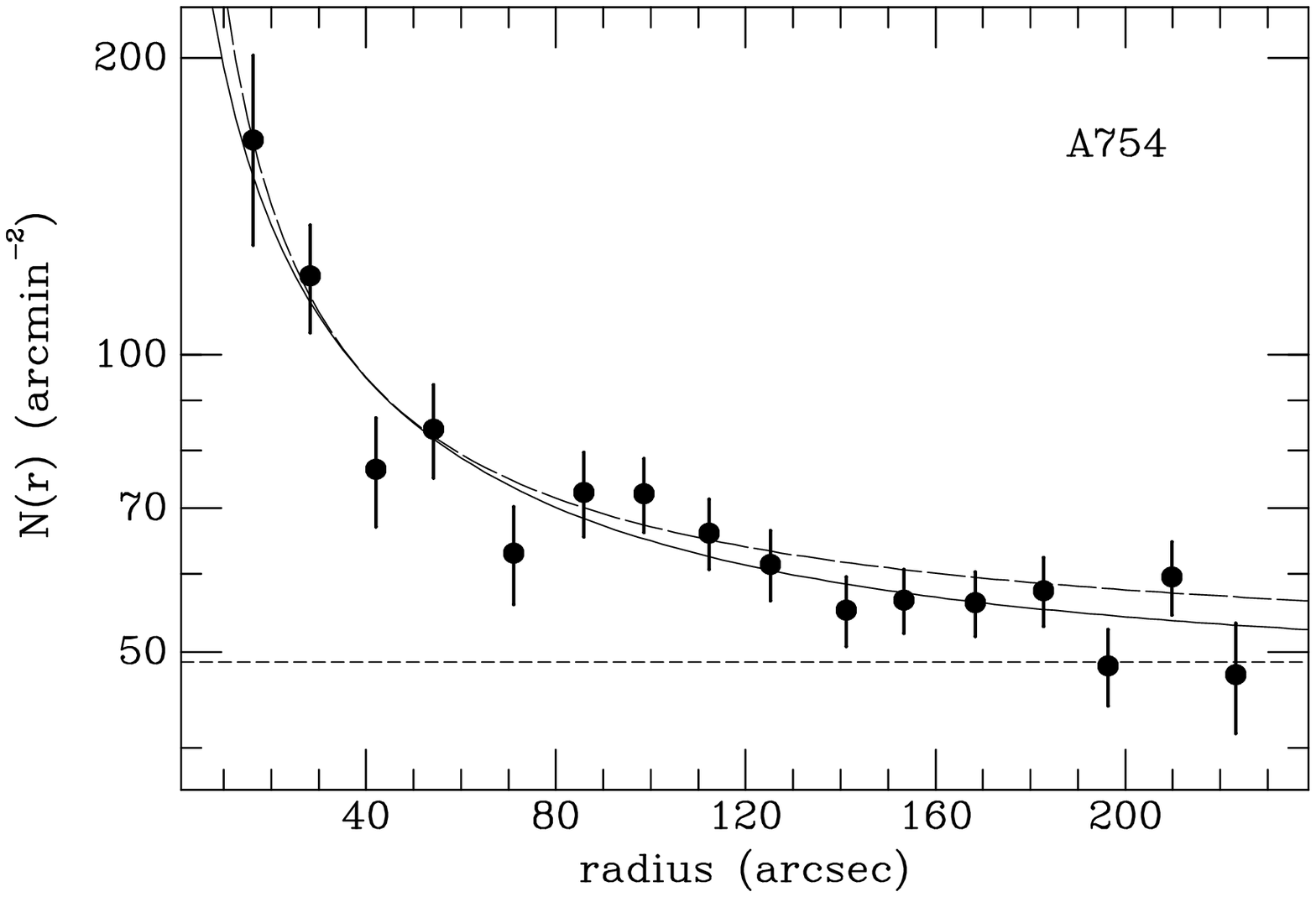}} 
\smallskip\caption{The incompleteness-corrected
number density of point sources is plotted as a function of radial
distance from the central galaxy in A754. 
The magnitude limits (corrected for Galactic extinction)
for the counts are given in Table~\ref{tab:nrad}.  The errorbars are
dominated by the counting statistics but include some contribution from the
uncertainty in the completeness.  The plot is log-linear in order to better
show the fractional uncertainty as a function of radius.  The solid curve
represents a 3-parameter de~Vaucouleurs $r^{{1/4}}$~law fit to all the
points shown; this fit was used to estimate the background point-source
count level, shown as a horizontal short-dashed line.  The background was
then subtracted, and a 2-parameter power law was fitted to the points within
a projected radius corresponding to $r<105\,$\hkpc\ (the uncertainty in the
background was included in doing this fit).  The long-dashed curve shows the
best-fit power-law model with the background added back in.}
\label{fig:a754rad}
\end{inlinefigure}

\begin{inlinefigure}\bigskip 
\centerline{\includegraphics[width=0.90\linewidth]{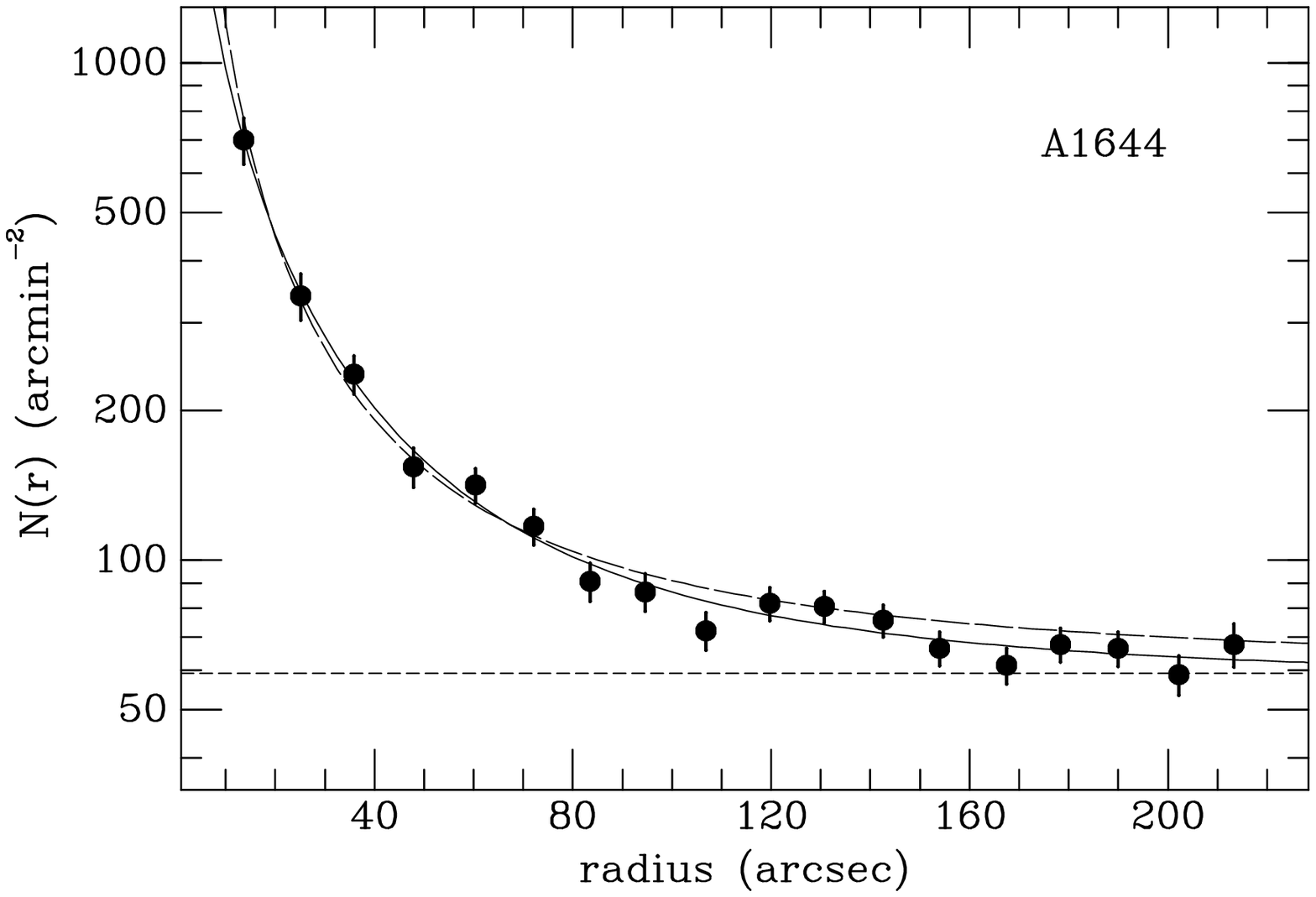}}\smallskip
\caption{Same as Fig.~\ref{fig:a754rad}, but for A1644.}
\label{fig:a1644rad}
\end{inlinefigure}

\begin{inlinefigure}\bigskip 
\centerline{\includegraphics[width=0.90\linewidth]{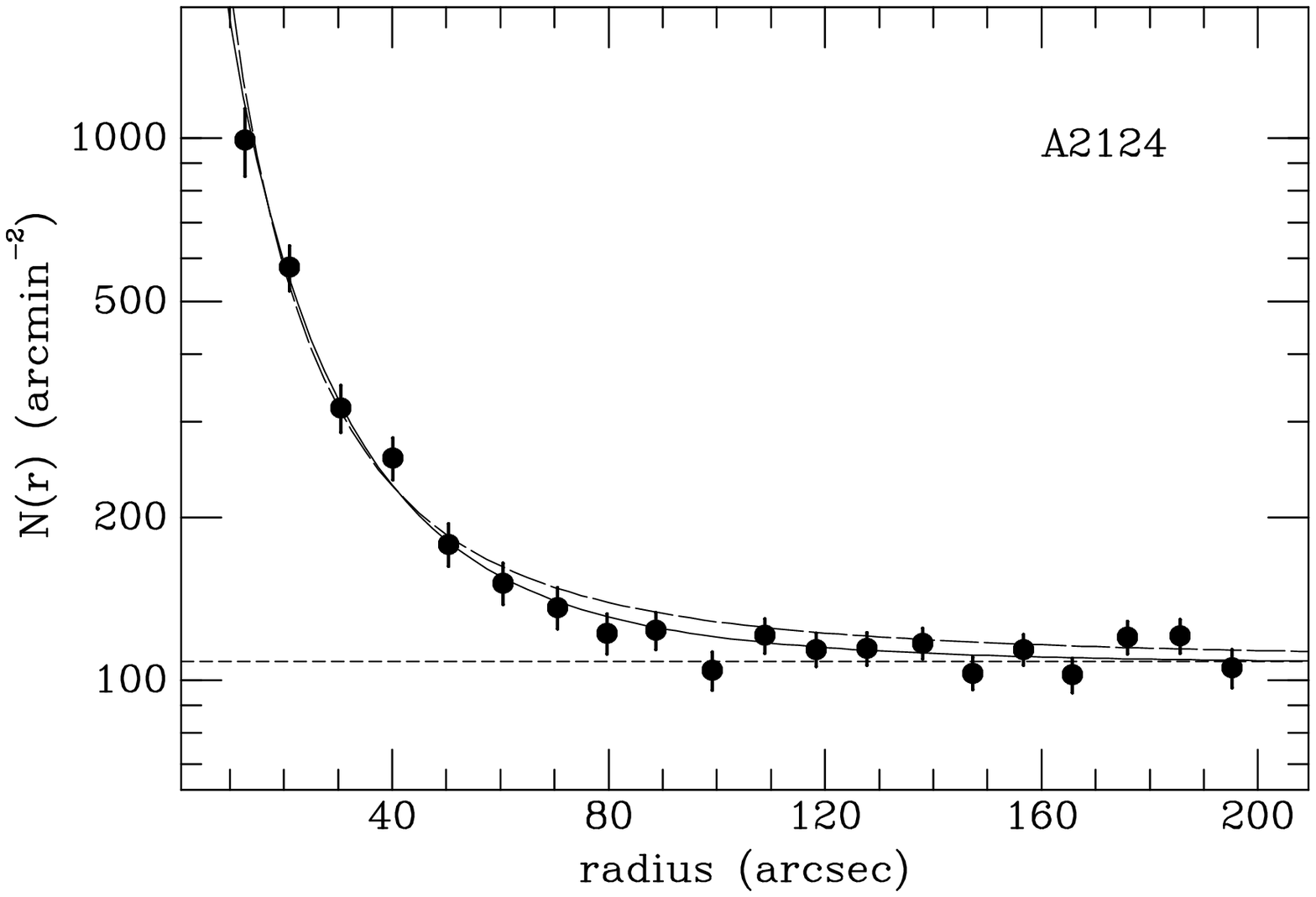}}\smallskip
\caption{Same as Fig.~\ref{fig:a754rad}, but for A2124.}
\label{fig:a2124rad}
\end{inlinefigure}

\begin{inlinefigure}\medskip 
\centerline{\includegraphics[width=0.90\linewidth]{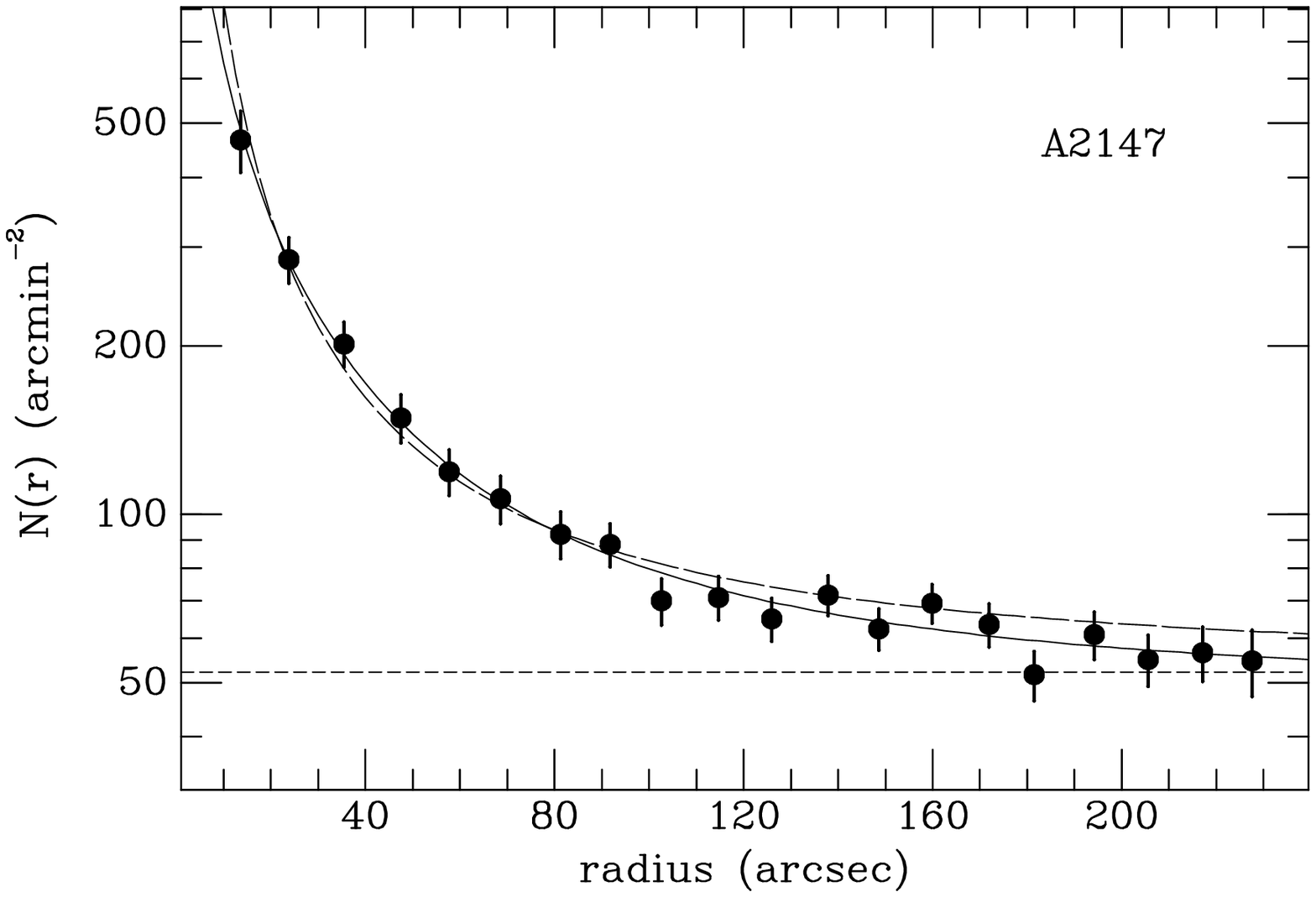}}\smallskip
\caption{Same as Fig.~\ref{fig:a754rad}, but for A2147.}
\label{fig:a2147rad}
\end{inlinefigure}

\begin{inlinefigure}\bigskip 
\centerline{\includegraphics[width=0.90\linewidth]{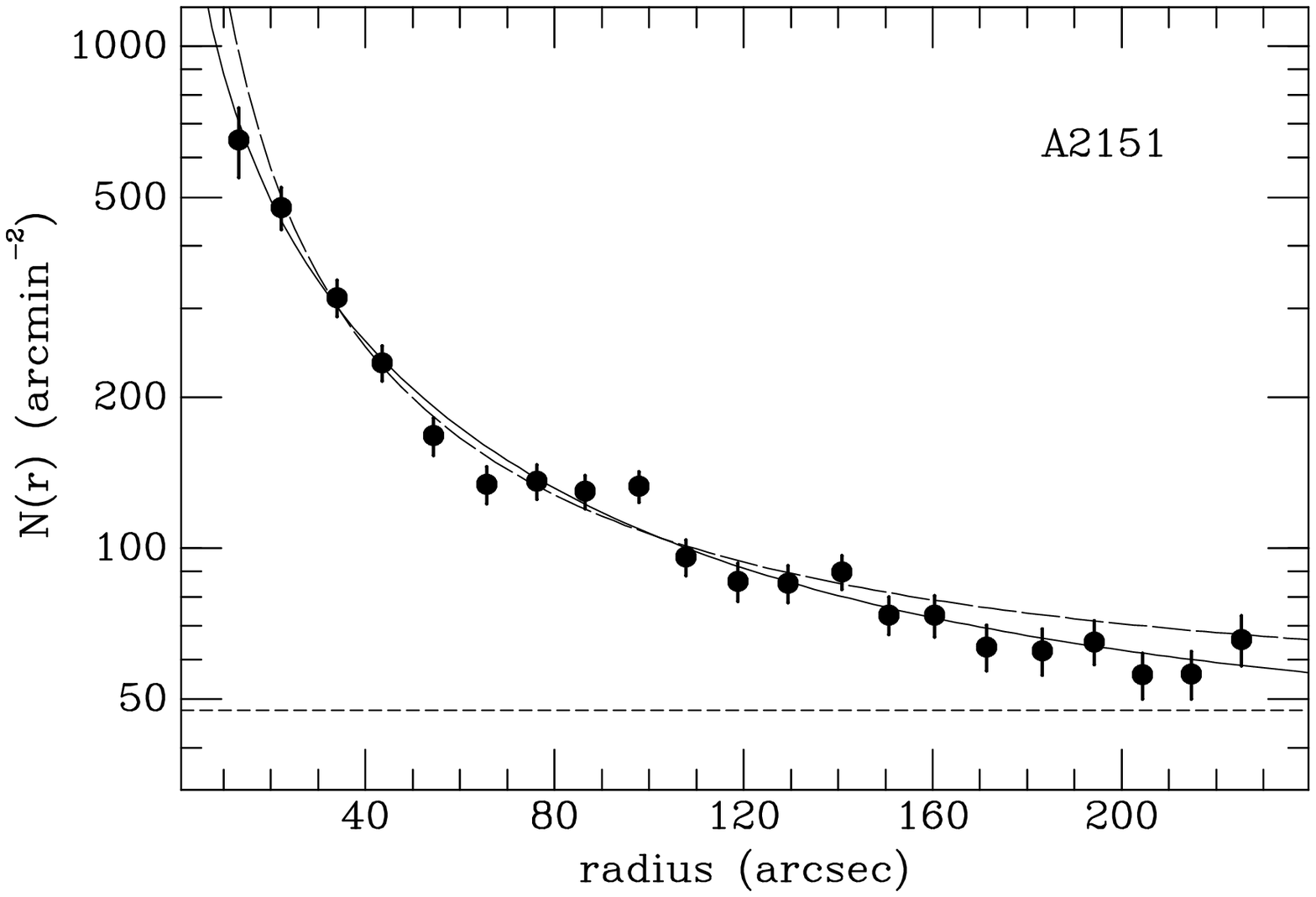}}\smallskip
\caption{Same as Fig.~\ref{fig:a754rad}, but for A2151.}
\label{fig:a2151rad}
\end{inlinefigure}

\begin{inlinefigure}\bigskip 
\centerline{\includegraphics[width=0.90\linewidth]{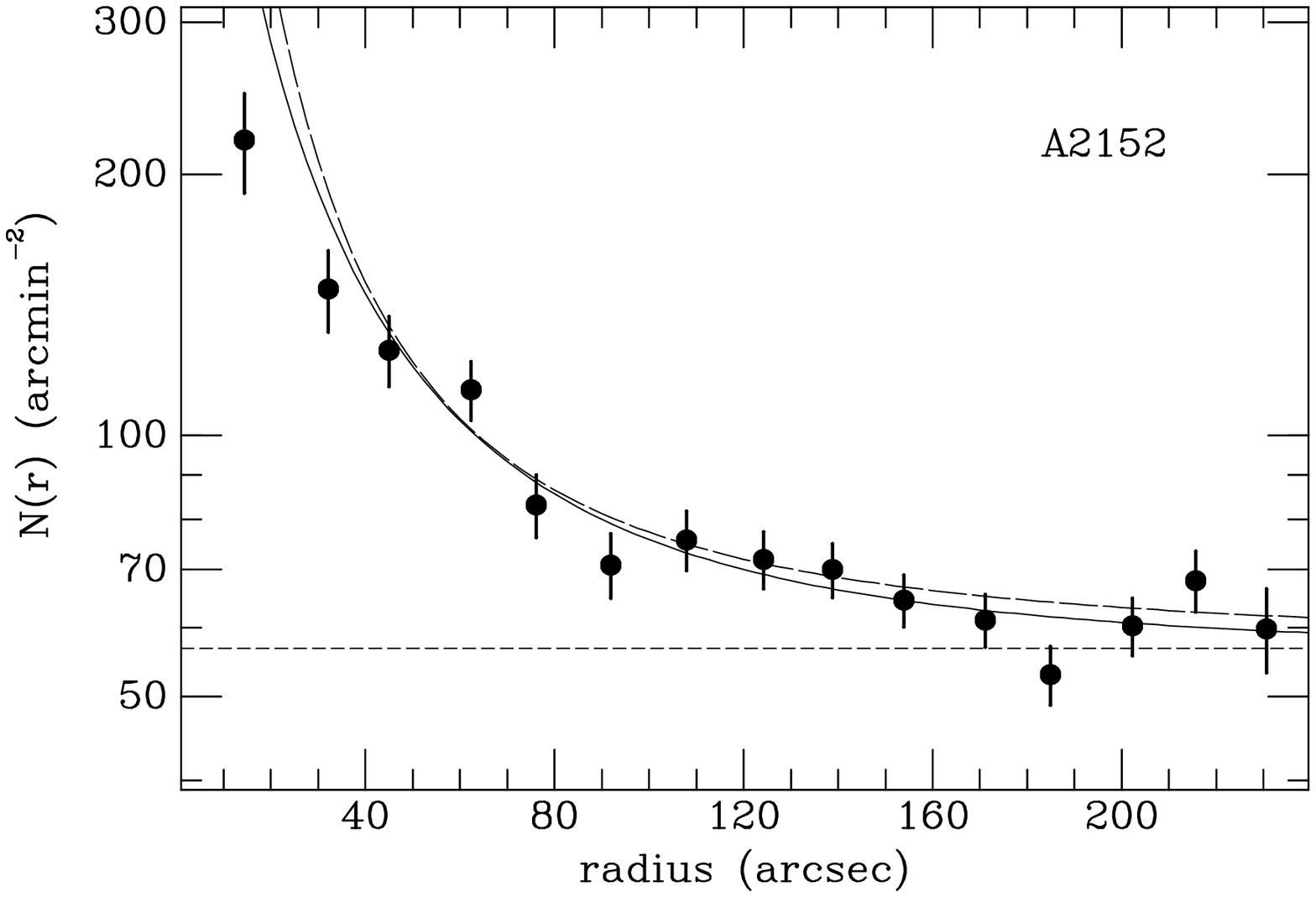}}\smallskip
\caption{Same as Fig.~\ref{fig:a754rad}, but for A2152.
Note that the luminous SRG is within a radius of 50\arcsec.}
\label{fig:a2152rad}\bigskip
\end{inlinefigure}

The radial number densities of GCs in galaxy halos are also frequently
parameterized as power laws of the form
\begin{equation}\label{eq:powlaw}
  N_{\rm GC}(r_p) \,=\, A_{0} \times r_p^{\alpha} \,,
\end{equation}
where the power-law exponent is typically $\alpha \approx -1.5$
for gE's, but with a range from $-1$ to nearly $-2.5$
(e.g., Harris 1991; Kissler-Patig 1997).
The background-subtracted surface densities of GCs for the present
sample of galaxies were fitted to power-law distributions of the
above form.  To aid in intercomparing the results, as well as to
provide a better comparison for studies of more nearby galaxies,
these power-law fits were done only within a radius corresponding
to a projected distance of 105~\hkpc\ from the BCG centers.
The fitted exponents are fairly insensitive to small changes in the 
magnitude or radial limits. 
These fits (with backgrounds added back in)
are also displayed in Figures~\ref{fig:a754rad}--\ref{fig:a2152rad},
and the parameters and magnitude range for each fit are listed in
Table~\ref{tab:nrad}.

It is important to bear in mind that the GC distributions shown in the
figures and characterized according to these simple radial forms actually
encompass multiple cluster galaxies in each case.  In particular for 
A754, A2151, and A2152 there are other bright galaxies in the fields with
fairly significant GC populations.  Specific frequencies are calculated
below for the secondary galaxies in these three clusters.  The galaxy
47\arcsec\ northwest of the BCG in A2152 is comparable in luminosity to the
BCG and has an obvious concentration of GCs; this is the main cause of
the flattening of the surface density distribution at small radii in this
cluster.

\subsection{Power Spectrum Measurements}

After all objects brighter than the chosen cutoff
magnitude $m_c$ are masked out of the image, the analysis proceeds
by a Fourier-space measurement of the variance in 
surface brightness due to objects fainter than $m_c$ remaining
in the image.  The power spectrum analysis is based on that used for
the surface brightness fluctuations (SBF) method of determining
galaxy distances (Tonry \& Schneider 1988).
A recent review of the SBF method is given by Blakeslee, Ajhar \&
Tonry (1999).  The measured power spectrum $P(k)$ is modeled as a
linear function of the ``expectation power spectrum'' $E(k)$,
which is the convolution of the \pf\ with the window function of the mask,
\begin{equation}
\label{eq:power}
P(k) \;=\; P_0 \times E(k) \,+\, P_1 \,.
\end{equation}
The ``white noise'' component $P_1$ is well determined by the high wavenumbers,
and so the problem becomes one of measuring the amplitude of the
``fluctuation power'' $P_0$.  BTM and Tonry \etal\ (1990) discuss
this measurement in detail and show example power spectra.

The $P_0$ fluctuation amplitude corresponds to the spatial variance in
intensity from sources that have been convolved with the image \pf.  
It has contributions from GCs, faint galaxies, and the stellar SBF.
The methods for estimating and subtracting the other contributions
to obtain \pgc, the fluctuation power due to GCs, were
developed by Blakeslee \& Tonry (1995) and have been discussed in more
detail by BTM.  The component due to the stellar SBF is estimated
by scaling the nearby gE measurements,
and the component due to faint galaxies is determined from
the background galaxy luminosity function measured in the image.
\pgc\ itself is directly proportional to the total surface density
of GCs in the image, according to:
\begin{eqnarray}
   P_{\rm GC} &=& \frac{N_0}{2} 
  \times 10^{0.8(m_1^*-m^0+0.4\siglf^2\ln[10])} \nonumber \\
  &\times&\erfc\left(m_c-m^0+0.8\siglf^2\ln[10] \over \sqrt2\,\siglf \right)
 \label{eq:pgc} 
\end{eqnarray}
(see Blakeslee \& Tonry 1995),
where $N_0$ is a normalization corresponding to the
total number of GCs per unit area,
\mo\ is the turnover magnitude of the GCLF, and
$\siglf$ is the Gaussian dispersion in magnitudes.
The proportionality factor thus depends on the GCLF, but
basically for elliptical galaxies
when $m_c$ is significantly brighter then \mo, then
$P_0$ is always dominated by GCs rather than by the stellar SBF.   
(This ratio of fluctuation powers
also depends on bandpass, as discussed by BTM.)

A new problem is created for the present data set by the variable 
\pf\ in the images.  The measured $P_0$ is sensitive to the adopted
\pf\ template used for $E(k)$.  To test the reliability and robustness
of the measured values, the fluctuation analysis was performed multiple
times in each field using all the available high signal-to-noise,
unsaturated stars in the region of the measurements as \pf\ templates
(at least 3 per field).  In addition, test reductions were done using
composite \pf\ stars made by subrastering and adding $\sim\,$20 individual
stars per frame, and then redone after spatially scaling the composite to
have the same FWHM as found in the region of interest.  In the end, one
reduction that gave results near the median was chosen per field,
and an allowance for an additional error of 6.25\% was made in accordance
with these tests to account for \pf\ mismatch.  This additional
uncertainty usually dominated the $P_0$ measurement error.

Table~\ref{tab:data} presents all the $P_0$ measurements and estimated
values of \pgc, and  \S\ref{sec:results} discusses how these are combined
with the count information to constrain the GCLFs and total population
sizes. First, however, the following section addresses one other correction
that must be applied for the current data sample.

\subsection{Gravitational Lensing Corrections}

The clusters in this sample are of sufficient masses and distances
that their gravity distorts the surface density of background sources.
For this reason, the background level close to the BCG (assumed to
be at the cluster center) may differ from that determined several
arcminutes farther out.  This effect is usually ignored in studies
of GC systems in clusters.  However, one of the sample clusters A2124
is known to possess a strongly lensed arc within a radius of 0\farcm5
of the BCG (Blakeslee \& Metzger 1999).  
Although it is the most distant in the present sample, A2124 is less
than a factor of two more distant than the nearest one, and it is not
especially massive.  Thus, lensing corrections were made for all
the clusters as described below.

A cluster lens both magnifies the background sources and dilutes
their surface density by some factor $A(r_p, z_\ell, z_s)$,
where $r_p$ is again the projected distance from the center of the mass
distribution (assumed round), $z_\ell$ is the redshift of the lensing
cluster, and $z_s$ is the redshift of the source.  
(Broadhurst \etal\ [1995] and
Trentham [1998] give fairly detailed discussions.) 
For lensing of {\em distant} sources by these clusters, 
$z_s \gg z_\ell \sim 0.05$, we can write the
effect of the cluster on the background as:
\begin{equation}\label{eq:lensN}
N_{\rm g}^\prime(m,r_p) \;=\; \frac{1}{A(r_p)}\, N_{\rm g}[m+2.5\log A(r_p)]
\end{equation}
where $N_{\rm g}(m)$ would be the apparent magnitude distribution of 
distant galaxies in the absence of lensing, and we have assumed
the sources lie at infinity and subsumed the cluster redshift 
into the definition of $A(r_p)$.
At the magnitudes of interest for the present analysis
$m_R \gta 25$, galaxies with luminosities $L \gta 0.1\,L^*$
would lie at redshifts $z_s \gta 1$.  (In the end, we make
a small correction for the finite redshift.)

As in BTM and Blakeslee \& Tonry (1995), we assume a power-law 
galaxy magnitude distribution of the form
\begin{equation}\label{eq:unlensN}
N_{\rm g}(m_R) \;=\; T_{\rm N} \times 10^{\,\beta\,m_R }\,,
\end{equation}
where a slope of $\beta = 0.35$ is adopted for the $R$-band counts
(Tyson 1988; Smail \etal\ 1995b; Hogg \etal\ 1997), consistent
with the faint magnitude distributions in the present data set
well away from the bright galaxies, and
the normalization $T_{\rm N}$ is derived from the data.
The lensing preserves the power-law form, and we can rewrite
Eq.\,(\ref{eq:lensN}) as
\begin{eqnarray}
N_{\rm g}^\prime(m_R,r_p) &=& [A(r_p)]^{-2.5(0.4-\beta)}\,
	 N_{\rm g}(m_R) \nonumber \\
  &\equiv& f_\ell(r)\, N_{\rm g}(m_R)\,,
\label{eq:lensplaw}
\end{eqnarray}
which defines the lensing correction factor $f_\ell(r)$.
Since $A(r_p) > 1$ and $\beta < 0.4$ here, the surface density at
a given apparent magnitude will be reduced by the lensing.

For a general lens, the magnification can be expressed as
(see Miralda-Escud\'e 1991)
\begin{equation}\label{eq:magfac}
A \,=\, \vert (1-\kappa)^2 - \gamma^2 \vert ^{-1}
\end{equation}
where $\kappa$ and $\gamma$ are the convergence and shear,
respectively.  The convergence is given by
\begin{equation}\label{eq:conv}
\kappa \; = \; {\Sigma(r_p)} 
	 \left(\frac{c^2}{4\pi G} \frac{D_s}{D_\ell D_{\ell s}}\right)^{-1}
  \, \equiv \; \frac{\Sigma(r_p)}{\Sigma_{\rm crit}} \,,
\end{equation}
where $\Sigma(r_p)$ is the surface mass density at a projected
radius $r_p$, and we have used the general definition for the
critical surface density 
$\Sigma_{\rm crit}$ in terms of the angular diameter distances from
observer to source $D_s$, from observer to lens $D_\ell$, and
from lens to source $D_{\ell s}$.  For an axisymmetric lens with the 
symmetry axis along the line of sight (i.e., an apparently round lens),
the shear is
\begin{equation}\label{eq:shear}
\gamma \; \equiv \; 
   \frac{\overline\Sigma(r_p) - \Sigma(r_p)}{\Sigma_{\rm crit}} \,,
\end{equation}
where $\overline\Sigma(r_p)$ is the mean surface density within $r_p$.
Thus, for an assumed mass model $\Sigma(r_p)$, one can correct the
background density using Eqs.~(\ref{eq:lensplaw})-(\ref{eq:shear}).

Ideally, one would use deep multicolor (or redshift) information 
to select out background objects, then use shape information for a 
weak lensing analysis to constrain the mass distribution, and then determine
the effect of this mass on the background surface density (extrapolated 
to fainter magnitudes).  
In practice, this would be extremely difficult because of
the fairly low surface masses involved, the severe
``contamination'' of the background by cluster GCs, the small number
of sources with measurable ellipticities, and the consequently coarse
resolution of the weak-lensing maps.  In any case, we lack 
multicolor or redshift information, and this level
of detail is not necessary for present purposes. 

Instead, the analysis explored a number of reasonable
mass distributions ranging from a singular isothermal sphere having the
cluster velocity dispersion, for which 
$\Sigma(r_p) = \case{1}{2}\overline\Sigma(r_p) = 
 (2Gr_p)^{-1}\sigma^2_{\rm cl}$\,,
to more complicated models such as a superposition of a Hernquist (1990) 
model\footnote[1]{A Hernquist model, or a Dehnen (1993) model
with $\gamma{\,=\,}1$, has a spatial mass density
$\rho = \frac{M_{\rm tot}}{2\pi}\frac{a}{r}(r+a)^{-3}$.  
It is similar to a Navarro, Frenk \& White
(1996) model in that $\rho \sim r^{-1}$ at small radii, but has the
advantage of being fully analytic.  For the lensing calculations, these
models were normalized by fixing the line of sight velocity dispersion 
to the tabulated values at $r_p = a$ and then $a$ was varied within limits.}
for the cluster and an isothermal sphere with $\sigma_{\rm g} = 300$ \kms\
for the central galaxy.  Isothermal models with cores were also explored.
While X-ray studies find cluster core radii of $r_c = 100\pm50$~\hkpc\
(e.g., Forman \& Jones 1999)
lensing studies usually find much smaller values in the
20--40~\hkpc\ range, or less 
(Mellier \etal\ 1993; Miralda-Escud\'e 1995;
Smail \etal\ 1995a; Tyson \etal\ 1998).  
For the case of A2124, Blakeslee \& Metzger (1999)
assumed the velocity dispersion in Table~\ref{tab:samp}
and found a best-fit isothermal lensing
model with $r_c \sim 10$ \hkpc.
We chose to model all of the clusters
as isothermal with a core radius $r_c = 20$~\hkpc.
This gives lensing factors between the extremes
of the singular isothermal and Hernquist models.
A comparison of the predictions from different mass models
suggests an uncertainty in the lensing factor (when integrated
over the radial ranges used here) that
is a third of the correction itself:
\begin{equation}\label{eq:lensunc}
\delta f_{\ell} \,=\, \case{1}{3}\,(1-f_{\ell})\,.
\end{equation}
This is included in the uncertainty estimates for the 
lensing-corrected background counts.

The lensing correction to the background density
gets as large as 27\% ($f_\ell = 0.73\pm0.09$) for the innermost
annulus (see the following section) of A2124.  However, with the
adopted $m_c = 26.1$ cutoff here, the pre-correction background is
only $\sim\,$10\% of the counts.  Thus, the lensing correction changes
the inferred number of GCs with $m_R < m_c$ by $\lta\,$3\%.
(In general, this correction has a slightly bigger effect on the
counts than on the power spectrum results because the slope of the
GC luminosity function at $m_c$ is steeper than that of the galaxies.)
Typical values of $f_\ell$ for this data set 
range from $\sim\,$0.84 for the innermost annulus to 0.98 for 
the outermost, but as the surface density of GCs falls with radius,
the fractional effect on the inferred number is roughly the same.
In the end, these corrections change the derived values for the
total numbers of GCs by $\lta\,$5\% and for the GCLF
widths by $\sim\,$0.01--0.02~mag.

\section{Results}\label{sec:results}

Table~\ref{tab:data} collects all the point source counts, fluctuation
measurements, and background estimates for all of the annular regions analyzed 
in each galaxy.  These regions each extend a factor of
two in radius, with the inner boundary of the innermost c1 region being
32~pix and the outer boundary of the outermost c4 region being 512~pix.
We follow the identical $\chi^2$ minimization procedure detailed by BTM.
We assume the usual Gaussian form for the GCLF and simultaneously
constrain the total numbers of GCs and the GCLF widths from the tabulated
measurements.  The expected GCLF mean (or turnover) magnitude \mo\ is
estimated for each cluster according to its velocity in the cosmic
microwave background (CMB) rest frame and the observed \mor\ 
in the Virgo cluster.  We use $m^0_R(\hbox{Virgo}) = 23.2$ 
(from Ferrarese \etal\ 1999 and the colors of Ajhar \etal\ 1994)
and assume an intrinsic scatter of 0.2~mag.
This calibration is 0.05~mag brighter than that
used by BTM, but we also revise the CMB velocity of Virgo to
$1280\pm70\,$\kms\ so that the calculated values of \mor\ (shown in
Table~\ref{tab:obs}) are consistent with those of BTM.  With the mean
Cepheid distance of $16.1\pm0.3\,$Mpc for 5 Virgo spirals (Ferrarese
\etal\ 1999), this gives $H_0 = 80$ \kmsMpc\ for calculating the
number of GCs per unit mass or luminosity. 
Changing the zero-point distance calibration would not affect the GCLF
results, which depend only on the relative distance with respect to
the Virgo cluster, and
it would change the results for \sn\ and \egc\ by only a constant
factor, so the conclusions regarding general trends would not change.
However, a systematic change in the distances with respect to Virgo
{\it would} systematically change the GCLF results, and we check for
this by comparing the results to the well-observed GCLF in Virgo.
Such a change has only a small effect on the values
for \sn\ and \egc, since the inferred number of GCs changes in
the same sense as the estimated masses and luminosities.
BTM and \S\ref{ssec:resgclf} below discuss these issues further.

Tables~\ref{tab:data} and~\ref{tab:Sn} (below) 
report measurements for three galaxies
besides the six BCGs in these clusters. A754-2 is 98\arcsec\ away at a
position angle $\hbox{PA} = 80^\circ$ from A754-1; A2151-2 is 92\arcsec\
away at $\hbox{PA} = 144^\circ$ from A2151-1; and A2152-2 is 47\arcsec\ 
away at $\hbox{PA} = 302^\circ$ from A2152-1.  The point source counts and
fluctuation analyses were done within annular regions centered around these
three galaxies as well.  Only the c1-c3 regions were analyzed for these; 
the circular region within 30\arcsec\ of A2152-1 was masked out during the
A2152-2 analysis (since these galaxies are so close together).  Otherwise,
very little attempt was made to protect against double counting of GCs
between the pairs of galaxies.  A2152-2 rivals A2152-1 in luminosity and 
was classified as the second~ranked galaxy (SRG) in this cluster by 
Postman \& Lauer (1995).  The other two ``secondary'' galaxies are 
significantly fainter than the respective primary cluster galaxies, but
they likewise had obvious concentrations of point sources around them.

\subsection{The GCLF, Biases, and Uncertainties}
\label{ssec:resgclf}

Table~\ref{tab:Sn} reports the results for the GCLF widths
\siglf\ and the specific frequencies $\sn^{40}$ and $\sn^{65}$
for different values of \siglf\ derived within metric apertures
of 40 and 65 kpc, respectively.  The uncertainties in \siglf\
include the effects of varying \mor\ within the $\pm0.22$ mag
uncertainty limits (derived from 0.2 mag intrinsic dispersion
and a 0.1 mag random uncertainty in the cluster Hubble velocities).
The weighted average (and the median) of the GCLF widths for all nine
galaxies is $\langle \siglf\rangle = 1.54\pm0.03$~mag with an rms
dispersion of 0.10~mag.  The six BCGs give the same weighted average
and dispersion with a median of 1.53 mag; if the A2124 BCG
(the most distant in the sample) is excluded, the average is
$\langle \siglf\rangle = 1.52\pm0.04$~mag with a dispersion of 0.08~mag.

BTM found $\langle \siglf\rangle = 1.45$~mag with a dispersion of
0.13~mag for their complete sample of 23 galaxies, but there was an
apparent bias such that the most poorly determined \siglf\ values
were also the largest.  When the 9 galaxies with the largest
uncertainties were excluded, the mean was 
$\langle \siglf\rangle = 1.43$~mag with a dispersion of 0.07~mag.
This result provided a good consistency check against M87, the central
galaxy in Virgo and the main calibrator for the GCLF in these
more distant clusters.  Whitmore \etal\ (1995) found $\siglf = 1.44$ mag
for their full sample of 1032 M87 GCs from Hubble Space Telescope
(HST) imaging.  The mean \siglf\ for the present sample of BCGs is 
$\sim0.1$~mag larger than these other values.
Although this result is only marginally significant at the
$\sim\,$2$\sigma$ level, and values of 
$\siglf \gta 1.55$ mag are not too uncommon in the literature 
(e.g., Madejsky \& Rabolli 1995; Elson \etal\ 1998),
we must consider the possibility that it is due to a 
systematic bias in the analysis.

If some systematic problem has caused \siglf\ to be overestimated for
the present sample, it is in the sense that the numbers derived from the
counts are too high by $\sim\,$25\% with respect to those derived from
the fluctuation analysis.  For instance, BTM discussed the possibility
that a population of unresolved dwarf galaxies clustering around the
larger galaxies could masquerade as GCs.  The implied properties of such
dwarfs and the simulations of Bassino \etal\ (1994) led BTM to reject
this as a serious problem. However, the potential for this problem
would be greater in these richer, more distant clusters; it may
be telling that the most distant cluster has the largest measured \siglf.
Future color information should help resolve this issue. 
BTM considered GCLF non-Gaussianity to be a bigger source of potential bias
for \siglf\ measurements.  If the GCLF has enhanced tails with respect
to a Gaussian, then the total number derived from the counts would be
overestimated.  Again, this problem is exacerbated when the cutoff
magnitude for the counts is much brighter than the GCLF turnover, e.g.,
$\mo{\,-\,}m_c \gta 2.5$ mag.  The small size of the current sample
precludes any definitive determination of the presence of this effect.
If it is true that the counts are simply overestimated due to
contamination and/or that the extrapolation from the counts is too high
because of non-Gaussianity, then the reported \sn\ values should be
decreased by 5-10\% (the fluctuation measurements carry more weight
because of their smaller uncertainties).

A bias in \siglf\ could also result from a systematic error in 
the $m_1^*$ photometric zero points or in the \mor\ values estimated
from the redshifts.
An error $\delta \mor$ or $\delta \mstar$ translates into a \siglf\ 
error $\delta\siglf \approx \case{1}{3}\delta \mor$ or 
$\delta\siglf \approx \case{1}{3}\delta m_1^*$.
Thus, the $m^0$ estimates or the $m_1^*$ zero~points would have to 
be wrong by 0.3~mag, in the sense that the data penetrate 0.3~mag 
further along the GCLF then we thought.
External photometric comparisons rule out a photometric error
of this size, and it is highly unlikely that the systematic error in 
\mor\ could be this large,
especially given the fact that BTM followed the same procedure. 
One change in the photometry for the present analysis is the use of
$R$-band extinctions from Schlegel \etal\ (1998); the 
Burstein \& Heiles (1984) extinctions for the program fields are less
by 0.04-0.12 mag.  Using the latter extinctions would decrease \siglf\
by as much as 0.03~mag in the mean.  However, it would
have little effect on the result for A2124, the one with largest \siglf,
since this field has the smallest change in its estimated extinction.

Alternatively, the zero points and \mor\ estimates could be
correct, but the \doph\ magnitudes could be systematically off.
The value of \siglf\ is much more sensitive to this:
an error $\delta m_c$ in the assumed cutoff magnitude
yields an error $\delta\siglf \approx 2\,\delta m_c$.
The reason is that Eq.\,(\ref{eq:pgc}) for the integrated number of GCs inferred
from the fluctuations is a function of both $(\mstar{-}\mor)$ and $(m_c{-}\mor)$, 
while the integrated number inferred from the bright counts is just a
function of $(m_c{-}\mor)$.  Changing the \mstar\ zero~point necessarily
changes $m_c$ as well, so the two sets of results change in similar ways.
However, an independent error in $m_c$ from the point-source photometry
affects the two results differently, and so makes a bigger
change in the value of \siglf.
The tests described in \S\ref{ssec:psource} for absolute offsets in the
aperture magnitudes, and for errors as a function of magnitude, indicate
that $\delta m_c < 0.03$ mag (in the one case with a significant
offset, a correction was made);
thus, it does not appear that this could be the whole explanation.
Perhaps several of the effects mentioned here contribute to a systematic
increase in the average \siglf, although, again, it is marginal.  
The possibility of intrinsic variations in the GCLF is discussed in
\S\ref{ssec:discgclf}.

\subsection{Specific Frequencies}

Blakeslee (1997) introduced the ``metric \sn'' as the value of \sn\
calculated within a fixed metric radius of 40~kpc; we will refer to this
quantity here as $\sn^{40}$.  This approach removes the need for large and
uncertain extrapolations to global values and reduces the risk of systematic
errors with distance.  Table~\ref{tab:Sn} reported values of $\sn^{40}$ for
the six BCGs and the three other ellipticals mentioned above.  The table
shows how $\sn^{40}$ decreases as \siglf\ increases.  Lacking clear evidence
for intrinsic variations of \siglf\ and in order to make the most homogeneous
comparisons, we take the approach of BTM and assume a single value of
$\siglf=1.45\pm0.05$~mag for the whole sample.

Figure~\ref{fig:sndisp} shows $\sn^{40}$ plotted against cluster velocity
dispersion $\sigma_{\rm cl}$ for the present sample and the BTM sample.
The BTM numbers have been decreased by 5\% to correct for the slightly
different distance scales used.  We have also switched to
$\sigma_{\rm cl} = 960$ \kms\ for the Coma cluster from Girardi \etal\ (1993)
rather then the extreme ``dense peak'' value of 1140~\kms\ from 
Zabludoff \etal\ (1993).  The correlation of \sn\ with $\sigma_{\rm cl}$
reported by Blakeslee (1997) and BTM persists, as the present sample of
galaxies falls in line at the high-density/high-\sn\ end of the relation.
The earlier sample also showed a strong increase in \sn\ with cluster
X-ray properties; Figure~\ref{fig:snxlum} shows $\sn^{40}$ plotted against
X-ray luminosity in the 0.5--4.5~keV band for the combined sample.  
The new data again fit in at the high end of the relation.
The quantitative relations minimizing the mean absolute deviations 
of the solid points in these two figures are given by:
\begin{eqnarray}
S_N^{40}  &=&  4.0 \,+\, 0.89 \times
\left(\sigma_{\rm cl} - 415 \over 100\right)\label{eq:snsig} \\
S_N^{40}  &=&  4.0 \,+\, 2.24\times(\log[L_{\rm X}] - 42.65) \,,
\end{eqnarray}
where $L_{\rm X}$ is in ergs~s$^{-1}$ and $\sigma_{\rm cl}$ is in \kms.

There is an important difference between the ``secondary'' galaxies
in the BTM and present samples, represented by the different open symbols in
Figures~\ref{fig:sndisp} and~\ref{fig:snxlum}.  The secondary galaxies
in the BTM sample (shown as open circles) are gE's with
luminosities similar to, and in some cases greater than, the primary
central galaxies in their respective clusters.  However, despite their
great luminosities, they are removed from the dynamical and X-ray
centers of the clusters.  The secondary galaxies in the present sample
(open stars) are lower luminosity ellipticals occupying the same
central fields as the BCGs in these clusters.  Thus, open circles
represent high-luminosity non-central galaxies, while open stars
represent low-luminosity central galaxies.  If the number of GCs
depends more on central location than on luminosity (as would be the
case if there is a significant population of intracluster GCs), then
one should expect the former galaxies to have lower specific
frequencies than the latter galaxies, which in turn should have
\sn\ values similar to those of the main central cluster galaxies.
This is precisely what these figures show.

\begin{inlinefigure}\bigskip\smallskip 
\centerline{\includegraphics[width=0.95\linewidth]{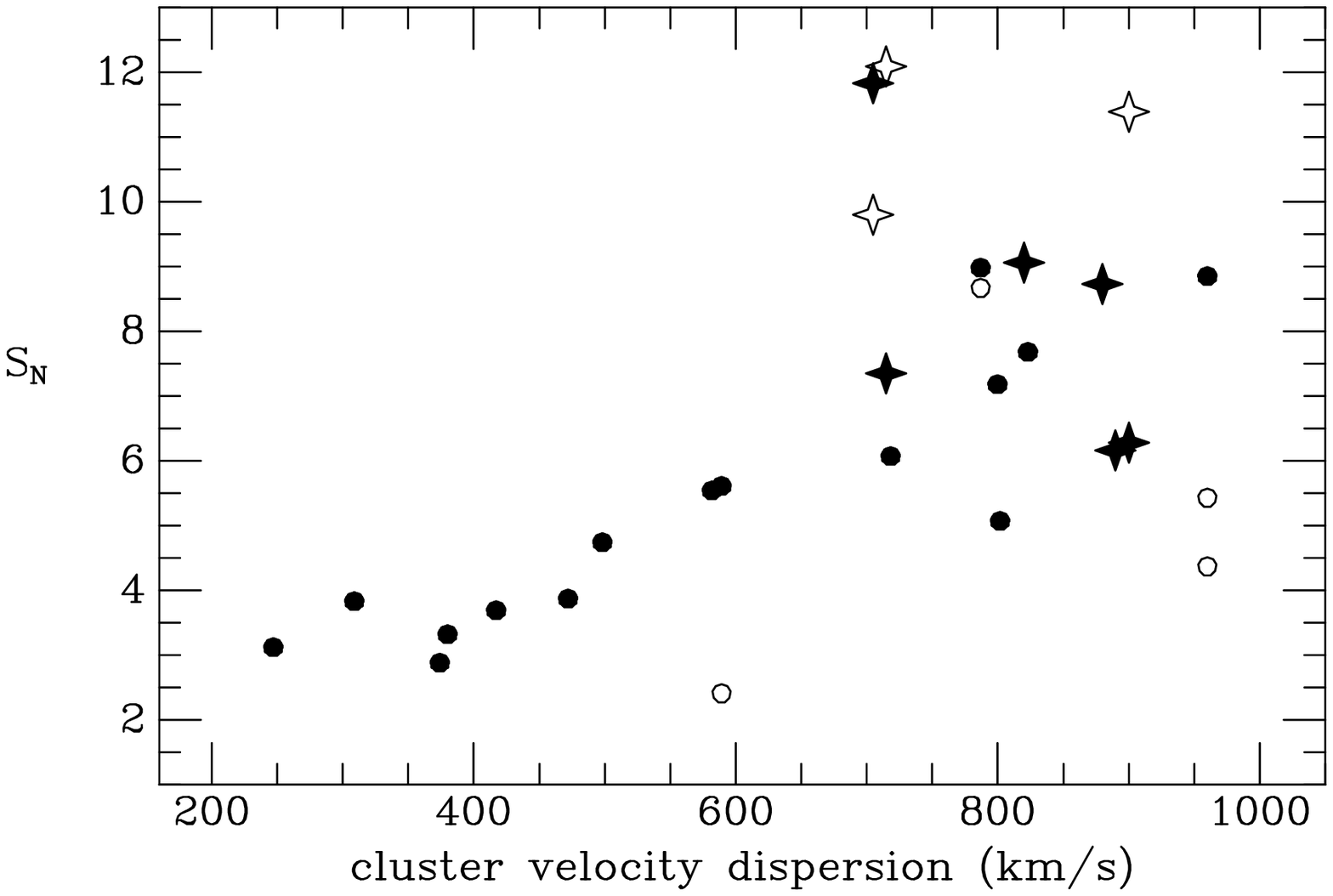}}\smallskip
\caption{
The correlation between GC specific frequency $S_N$ and the velocity dispersion
of the galaxies in the host cluster.  Solid and open circles respectively 
represent the central and non-central (``secondary'') cluster galaxies from
the BTM sample.  The solid four-pointed stars show the results for the main
central cluster galaxies in the present sample, and the open stars are for
other ellipticals in three of these same central fields.  See text for details.
\label{fig:sndisp}}
\end{inlinefigure}

\begin{inlinefigure}\bigskip 
\centerline{\includegraphics[width=0.95\linewidth]{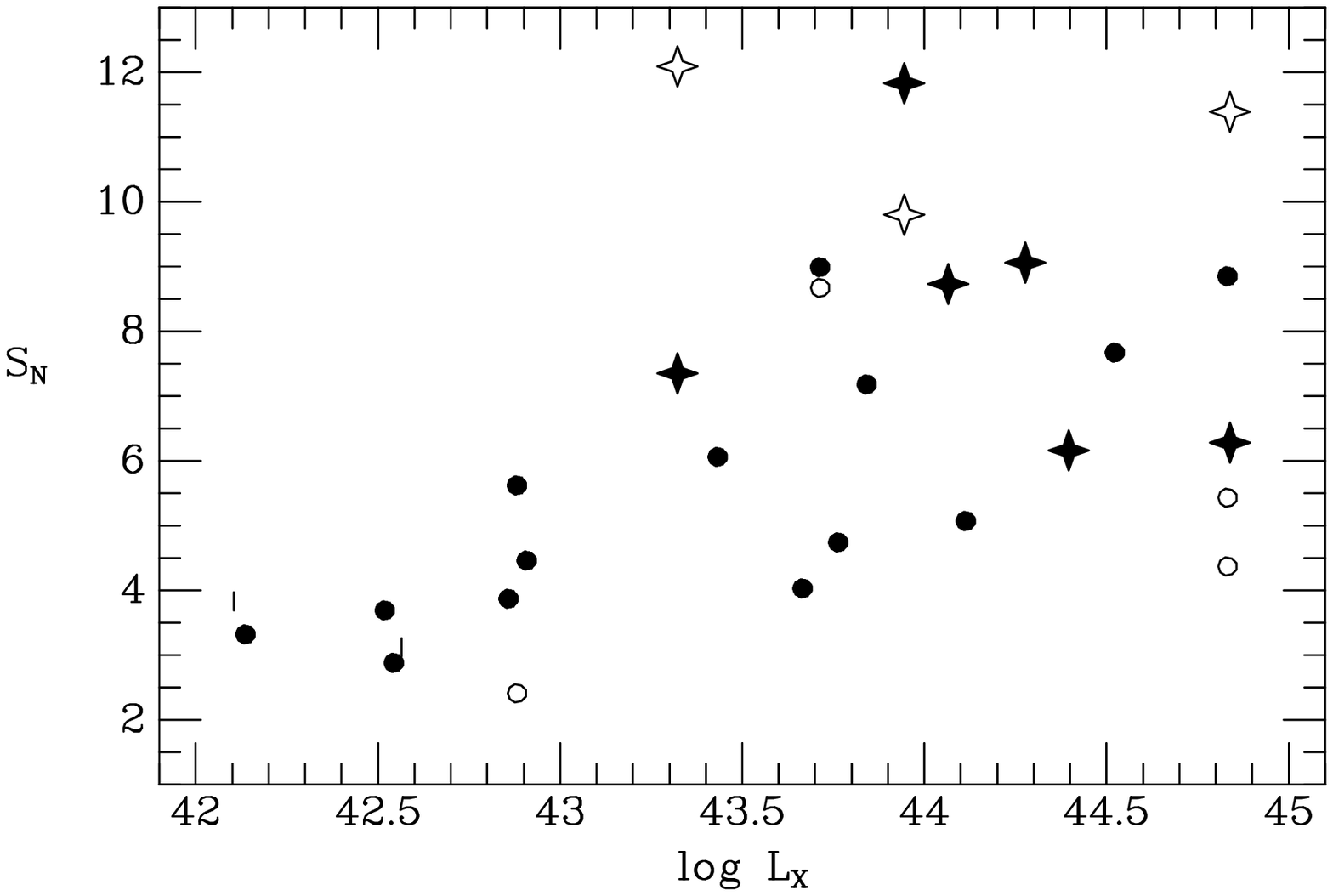}}\smallskip
\caption{
The correlation between GC specific frequency $S_N$ and the X-ray luminosity
of the host galaxy cluster.  Symbols are as in Fig.~\ref{fig:sndisp},
except for the short vertical lines which represent BTM galaxies in
clusters with only upper limits on X-ray luminosity.
\label{fig:snxlum}}
\end{inlinefigure}\medskip

The larger area of the LRIS field of view and the greater distances of
the present sample of galaxies allows \sn\ values to be measured over a
larger range in radius.  The last column of Table~\ref{tab:Sn} lists
values for $\sn^{65}$, the value of \sn\ within a metric radius of 65~kpc.
These numbers provide better approximations to the global \sn\ values 
of the central cluster galaxies; the tabulated errors include 
uncertainties of $\pm0.22$~mag in \mor\ and $\pm0.05$ mag in \siglf.
The general increase in \sn\ between 40 and 65 kpc indicates that the
GC distributions are generally more extended than the halo light.
There is nearly a factor of two variation in \sn\ among these galaxies,
and although this sample as a whole fits in nicely at the high end
of the \sn-density correlations, there is no such correlation 
apparent within this sample itself.  Three of the BCGs have 
$\sn^{65}=9$--10; the two most luminous ones have $\sn^{65}\sim7$,
and the A2151 BCG has $\sn^{65}=12.5$.  A2151 is only remarkable
in being a fairly irregular, high spiral-fraction cluster; thus,
it may be relatively young and dynamically unevolved.
Possible implications of this observation are discussed
in \S\ref{ssec:evolution}.

\subsection{GCs per Unit Mass}

Motivated by the increase in GC number with cluster density and the
relative constancy of BCG luminosity which makes \sn\ depend on
density, BTM defined the quantity \egc\ as the number of GCs per unit
mass: $\egc = N_{\rm GC}/M_{\rm cl}$, where $M_{\rm cl}$ here is the
total mass in units of $10^9\,M_\odot$ within the same projected radius
in which $N_{\rm GC}$ is measured.  The procedure of using projected
numbers and masses is analogous to calculating \sn\ from projected
luminosity and number; Harris \etal\ (1998) take a different approach.
The BTM \egc\ parameter is similar to the $T$ parameter defined by Zepf
\& Ashman (1993) as the number of GCs per unit mass for isolated galaxies.

Figure~\ref{fig:egc} shows \egc\ calculated for the combined BTM
plus present sample using two different cluster mass models.
The top panel uses a singular isothermal model, for which
$M({<}r_p) = ({{\pi \over G}})\sigma^2_{\rm cl}\,r_p$, and the lower
panel uses the flat (projected) core model of BTM, for which 
$M({<}r_p) \sim \sigma^{1.4}_{\rm cl}\,r_p^2$.  Cohen \& Ryzhov (1997)
used the dynamics of the M87 GCs and found that the mass profile in
Virgo was between these two cases.  The masses are calculated between
projected radii of $\sim4$~kpc and 40~kpc, the range appropriate to
the GC number.  Thus, the ``singular'' isothermal model could be made
to flatten within 4~kpc without changing the results.
Because the present calculation now more properly excludes this central
region, and because of the slightly longer distance scale used here, the
BTM data points shown in the lower panel of the present
Figure~\ref{fig:egc} are 6\% higher than they were in BTM's own
Figure~15, except for the Coma point, which is $\sim30$\% higher because
it also uses a different velocity dispersion.

The isothermal model of the top panel gives $\egc\sim0.5$ but probably
overestimates the true mass; the flat core model of the lower panel
gives $\egc\sim0.9$ but likely underestimates the mass because it
makes no account for a mass peak near the central galaxy.  In either
case, the scatter is about 30\% and there is little trend
left with $\sigma_{\rm cl}$.  
The highest point in both panels is A2151, which has the smallest
velocity dispersion in the present sample; a dispersion more in line
with its richness would lower its \egc.  Without A2151, 
the scatter in \egc\ for the flat core model would be closer
to 20\%, while the isothermal model would show a marginal decline in \egc\ 
with $\sigma_{\rm cl}$.  However, the isothermal model gives more
consistent results with radius, since its projected mass density
goes as $r_p^{-1}$ and the GC surface densities in
Figures~\ref{fig:a754rad}--\ref{fig:a2152rad} go as $\sim r_p^{-1.5}$.
Table~\ref{tab:egc} lists the values of \egc\
calculated within 40 and 65 kpc for each of the two mass models.
The calculated \egc\ drops by about 40\% for the flat core model
but only by about 10\% for the isothermal model.  
Some drop would be expected if the total GC system is a combination 
of cluster and galaxy GC populations.

The new results are in good accord with those of BTM, who proposed that
the apparent universality of \egc\ (at least when calculated within a
fixed metric radius) reflects a universal GC formation efficiency in the
dense primordial environments that have become today's galaxy clusters.
Thus, there remains no cause for invoking a density-dependent GC
formation efficiency.
The following section considers what light the present data shed on the
formation, evolution, and luminosity functions of GC systems in clusters.

\begin{inlinefigure}\bigskip 
\centerline{\includegraphics[width=0.95\linewidth]{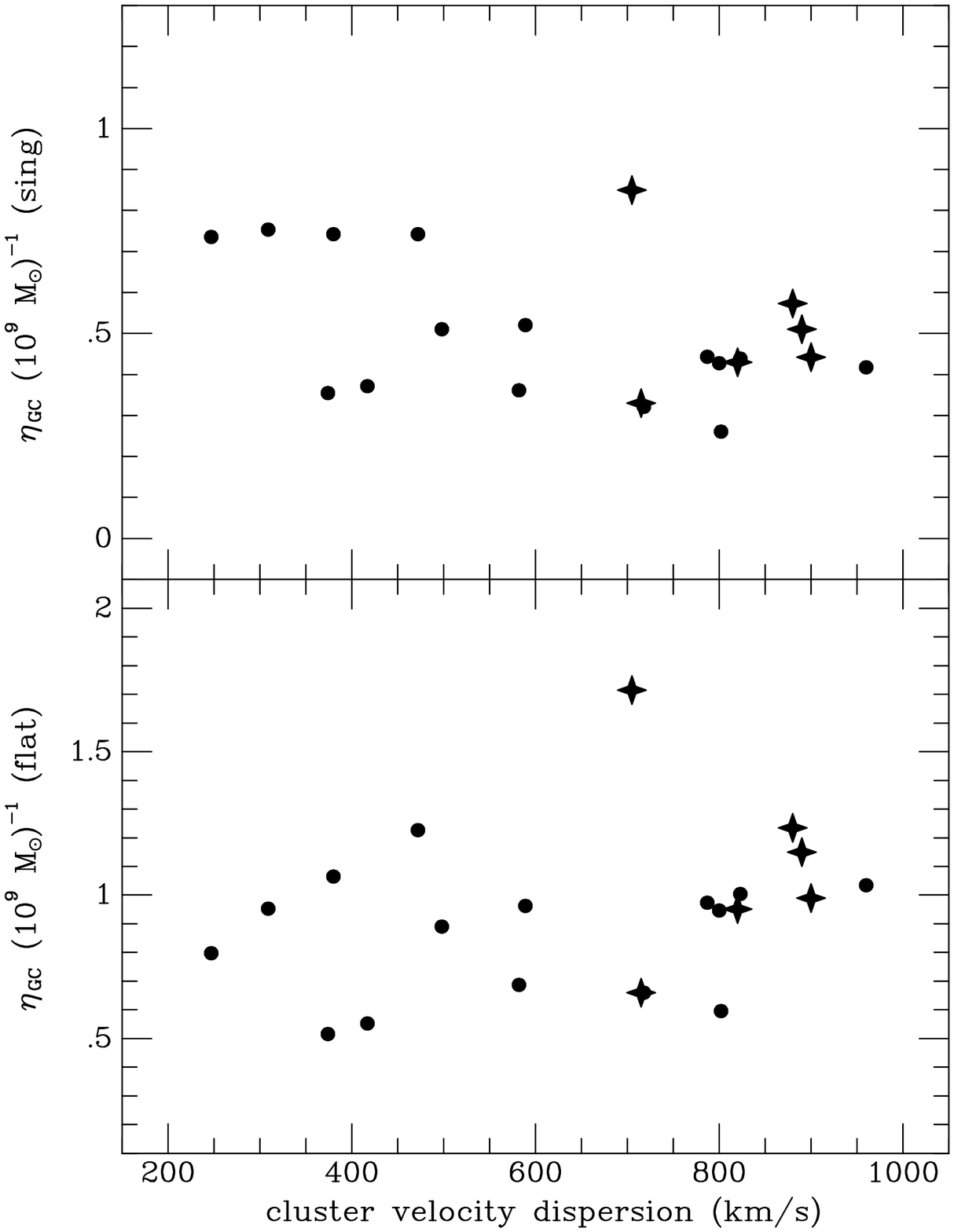}}\smallskip
\caption{
The number of GCs per unit mass $\eta_{_{GC}}$ is plotted against
cluster velocity dispersion for the central cluster galaxies in the BTM 
(solid circles) and present (solid stars) samples.  Both a singular
isothermal model (top) and the flat core model of BTM (bottom) are used
for estimating the masses.  The average values in the two panels are
$0.50\pm0.16$ and $0.93\pm0.27$ GC per $10^9\,M_\odot$, respectively.
\label{fig:egc}}
\end{inlinefigure}

\section{Discussion}

As their numbers scale with overall properties of the galaxy clusters, the
GCs of the central cluster galaxy might more correctly be called ``cluster
globulars'' (or cluster GCs) as opposed to galactic ones (cf. West \etal\
1995; BTM). Of course, it may be more appropriate to break the total
population down into cluster and galactic components (e.g., C\^ot\'e
\etal\ 1998).  This does not suppose that the putative cluster GC
population formed outside of galaxies, as it is difficult to imagine that
the very dense regions where the GCs must have originated did not
eventually become galaxies.  However, these galaxies would be ``cluster
galaxies''; thus, all of their associated GCs would have the potential to
become ``cluster globulars,'' whether through early escape during the
relaxation of the proto-galaxy, by later stripping during interactions,
or by falling into the center of the potential well during cluster collapse
to help form the dominant galaxy there.

\subsection{Universal Formation Efficiencies}
\label{ssec:effic}

McLaughlin (1999b) contends that his view of the universal GC 
formation efficiency is ``quite different'' from that of BTM,
and thus different from the one given here.  His main point of
contention is that BTM estimated efficiencies in terms of the 
number of GCs per unit total mass, instead of using only the 
combined mass of the stars and gas, i.e., the ``baryonic mass.''
BTM used the most convenient observable at hand to estimate
masses, namely the cluster velocity dispersion.  No mention 
of the role of nonbaryonic dark matter was made in that work,
apart from the statement that this approach obviated the need
for ill-constrained assumptions such as dark matter biasing or a 
density-dependent efficiency.

McLaughlin (1999b) is incorrect in stating that the BTM efficiencies
were computed using the total masses of the ``{\it entire} galaxy clusters''
(italics his).   Rather, the masses were estimated within 
the identical projected radii in which the numbers of GCs were measured
(since we can only observe clusters in projection).  
If the GCs, which are known to scale with cluster properties,
roughly follow the mass distribution, then the results computed
within projected radii will be similar to those found within
three-dimensional radii.  In addition, the universal efficiencies
proffered by BTM and McLaughlin will be equivalent, modulo a scale
factor equal to the ratio of the masses used.
In a different work, McLaughlin (1999a) concludes that the 
mass distribution of the hot gas follows
the overall dark matter distribution, at least at large radii.
He quotes a ratio of the gas mass to dark matter mass in Virgo of
$\rho_{\rm gas}/\rho_{\sc dm} \sim 0.035$.
Thus, to convert between the different efficiencies, 
we need now only to estimate the ratio of the stellar and gas masses
over the range in projected radii for which the GC surface densities
were measured, i.e., from $\sim\,$4~to 40~kpc.

For the gas, we adopt the Hernquist (1990) mass model 
from McLaughlin (1999a,b) to obtain a projected mass
$M_{\rm gas} = 1.77\times10^{11}\,\msun$
over this range in projected radius.
To estimate the stellar mass, we use a Hernquist model
having the same effective radius $R_{\rm e} = 7.0$ kpc
as the photometry used by McLaughlin (de~Vaucouleurs \& Nieto 1978),
which extends out to $\sim\,$100 kpc.  We normalize the model
such that the spatial mass density of the stars $\rho_{\star}(r)$
equals that of the gas $\rho_{\rm gas}(r)$ at $r=42$ kpc
(McLaughlin 1999b). This yields 
$M_{\star} = 4.28\times10^{11}\,\msun$ for this projected radial range;
the appropriate projected mass ratio is then
$M_{\star}/M_{\rm gas} = 2.4$.
The identical value is obtained if one instead uses the 
McLaughlin (1999a) mass model for the stellar distribution
(a $\gamma{\,=\,}1.33$ Dehnen model).
Thus the appropriate ratio of total to baryonic projected
mass is
\begin{eqnarray}
M_{\rm tot}/(M_{\rm gas}{+}M_{\star}) &=&
(M_{\rm gas}{+}M_{\star}{+}M_{\rm DM})/(M_{\rm gas}{+}M_{\star}) \nonumber \\
 &\approx& (1 + 2.4 + 0.035^{-1})/(1+2.4)\; \nonumber \\ 
 &\approx& 9.4\;.	\label{eq:masses} 
\end{eqnarray}

The average of the values for the GC mass rate \egc\ from the
two panels of Figure~\ref{fig:egc} is
$$\egc \,=\; 0.71 \pm 0.22 \;\;\hbox{GC~$M_\odot^{-9}$} \,.$$  
This is similar to the value quoted by BTM, so we will work with it. 
To convert this to a dimensionless efficiency, we adopt a mean
individual GC mass of $\langle m_{\sc GC} \rangle = 2.4 \times
10^5\,\msun$, as in McLaughlin (1999b).  The inferred GC formation
efficiency per unit total mass is then 
$\epsilon_{\rm tot} = (1.71 \pm 0.53) \times 10^{-4}$. 
Using Eq.\,(\ref{eq:masses}) to convert this to an
efficiency per unit baryonic mass, we obtain $\epsilon_{\rm b} = 0.0016
\pm 0.0005$.  The errorbar here should be considered a rough lower limit
on the uncertainty, which may be nearer to $\sim\,$50\%, given the
fairly crude mass models employed.  In any case, this result is close to
McLaughlin's proposed universal value of $\epsilon_{\rm b} =
0.0026 \pm 0.0005$.

McLaughlin (1999a) remarks on the similarity of the total
stellar and gas masses which makes the baryon fraction
$M_{\rm b}/M_{\rm DM} \sim 0.07$. He also
notes that the dark matter distribution
is not constrained for $r\lta15$ kpc. 
In the model of Harris \etal\ (1998), the strong galactic
winds that removed the star-forming gas from the proto-cD
would have left the dark matter distribution intact, as
this material is supposed to interact only gravitationally.
Taking the view of Waxman \& Miralda-Escud\'e (1995), Navarro \etal\ (1996),
and Arag\'on-Salamanca, Baugh, \& Kauffmann (1998)
that the dark matter is {\it more} centrally concentrated
than the hot gas, and thus more closely follows the overall
baryonic mass distribution, we find
$\epsilon_{b} \approx (1+0.07^{-1})\times1.71{\times}10^{-4} = 0.0026$,
fortuitously identical to the McLaughlin value.

Given these considerations, McLaughlin's universal GC formation
efficiency appears identical in every palpable way to that of BTM, who
presented the initial observational evidence for it and made a rough
first estimate of its value.  (The idea itself goes back at least twenty
years to Harris \& Petrie [1978].)  McLaughlin's work has substantially
refined the idea and allowed for significant progress on the physical
mechanisms underlying the observations.

\subsection{GC Color Distributions}

Up to now, this discussion has ignored the vast wealth of recent data on
the color distributions of GC populations (e.g., Ostrov \etal\ 1993;
Whitmore \etal\ 1995; Geisler \etal\ 1996).  The complex, often bimodal GC
color distributions of gE galaxies must surely be important clues
bespeaking a complex enrichment history.  
The situation is simpler for dwarf ellipticals, which appear to
have only metal poor GCs (see C\^ot\'e \etal\ 1998).  In hierarchical
formation models, giant galaxies are constructed (at early times for dense
environments) through chaotic merging of numerous smaller subunits that
must have resembled dwarf galaxies (e.g., Searle \& Zinn 1978).  To be
consistent with observations, the gE progenitors must have formed their
high metallicity GCs during this process; several lines of argument
suggest this is possible (see Ashman \& Zepf 1992). 
The final GC metallicity distribution will be broader and more metal rich 
than that of the dwarf progenitors, but in a continuous
hierarchical collapse, there is no {\it a priori} reason to expect 
distinct metallicity peaks or a change in \sn\ to occur. 
C\^ot\'e \etal\ (1998) have demonstrated using
realistic galaxy luminosity functions that later dissipationless mergers
can result in gE's having bimodal GC distributions, again without any
change in \sn.

In any case, the similarity in the GC color distributions of cD galaxies
and more ordinary ellipticals (Forbes \etal\ 1997) reinforces the
conclusion of BTM that there is nothing special, or unusual, about the GCs
around central cluster galaxies.  For instance, the GC color distributions
in M87 and M49 in Virgo are both bimodal in appearance, despite the fact
that M87's \sn\ is about 3 times higher.
Kissler-Patig \etal\ (1999) pointed out that the similarity between the GC
color distributions of the Fornax cD NGC 1399 and the other Fornax
ellipticals is expected, despite NGC 1399 having an \sn\ twice that of 
any of the others, if \sn\ increases by a simple transfer of GCs through tidal
stripping.  However, stripping scenarios in virialized clusters
have unsolved problems with time scale (Muzzio 1987), and it is not 
at all clear that \sn\ should increase, since both stars and GCs will be added
(e.g. Harris 1991), although it might (e.g., C\^ot\'e \etal\ 1998).

\subsection{BCG Formation and the Evolution of $S_N$}
\label{ssec:evolution}

Since both the GC color distributions and numbers per unit mass are normal
in high-\sn\ galaxies, the question is one of the missing
light: why do these galaxies lack the halo light necessary to bring their
specific frequencies and mass-to-light ratios down to normal levels?  
Motivated by the ``standard candle'' aspect of BCGs (e.g., Postman \& 
Lauer 1995), BTM suggested that the collapse of the galaxy cluster tidally
heated and removed the gas from the central galaxy and halted star 
formation there.  Improvements in numerical modeling of BCG formation can
help to refine this idea.

Recent simulations indicate that giant central galaxies form on
relatively short time scales during the initial collapse of the cluster.
Garijo, Athanassoula, \& Garc\'ia-G\'omez (1997) simulated the formation
of central galaxies in clusters with a variety of different initial conditions.
They observed rapid growth of a central dominant galaxy and the early
formation of an extended halo of debris.  The cD obtained
a great mass within 1~Gyr for their higher velocity dispersion clusters.
Dubinski (1998) studied the collapse and evolution of a 550~\kms\
dispersion cluster in a hierarchical formation simulation starting at
$z{\,=\,}2$ and found that the most massive galaxies all merged together to
form one central object in less than 3~Gyr (by $z{\,=\,}0.8$).  Some
other galaxies were subsequently accreted, but there were no more large
mergers in the cluster center after $z{\,=\,}0.4$.

Unfortunately, the simulations do not yet include modeling of the
gas dynamics and star formation in the collapse.  Starbursts would
likely occur during the rapid series of mergers, but arguing that
GCs must have formed at a higher than normal rate in some clusters
(relative to the total number of stars formed) does not solve the problem,
which results from ``missing light,'' or ``luminosity saturation,''
not ``excess globulars.'' 
If the simulations are correct that massive central galaxies
form quickly, and more quickly in the more massive protoclusters, then the
simple assumptions that, (1)~most GCs are older than most stars, and 
(2)~the central galaxy quickly loses the cool gas necessary to form stars,
are sufficient to explain the observations.  
The first of these assumptions is true in our own Galaxy, and spectroscopy
of a large sample of M87 GCs indicates that they are uniformly old as well
(Cohen \etal\ 1998).  The second assumption is true today except in rare
cases ($\sim\,$4\%, according to Lauer \& Postman 1994).
Arag\'on-Salamanca \etal\ (1998) conclude from BCG colors
out to $z{\,\sim\,}1$ that no significant star formation
has taken place in BCGs since that epoch;
they suggest that the gas is less centrally concentrated than the
dark matter, and hence unable to cool efficiently enough to form~stars.  

Both tidal heating (BTM) and strong, starburst-driven galactic winds
(Harris \etal\ 1998) during the cluster-collapse/BCG-formation epoch
have been offered as means for removing gas from the BCG.  
The faster formation of BCGs in denser (simulated) environments 
would drive the gas out sooner.
As an example, assume that the GCs formed first on a fixed
short time scale, so that their total number is proportional to
the mass in an isothermal model:
$N_{\rm GC} \sim M_{\rm cl} \sim \sigcl^2$.
Also assume that the amount of material for forming non-GC stars
has this dependence, but that the time $t_\star$ 
available for forming the stars 
is inversely proportional to the velocity dispersion:
$N_{\star} \sim M_{\rm cl} t_\star \sim \sigcl$. Thus, 
\begin{equation}
\sn\sim N_{\rm GC}/N_{\star} \sim \sigcl\,,
\end{equation}
and we reproduce a linear dependence as in
the empirical relation given by Eq.\,(\ref{eq:snsig}).
A more elaborate scenario awaits further advances in the simulations,
but an analogy can be drawn with the situation in the Milky Way.
If we assume that 90\% of Galactic stars formed after the GCs did
and use the {\it current} mass-to-light ratio $(M_\star/L_V)_{0}$ 
of the Galactic stellar population to compute $\sn^\prime(t)$, 
a ``corrected'' specific frequency at that early time, then
\begin{equation}
S_N^\prime(t) \,=\, {{N_{\rm GC}} \over {M_\star(t)/(M_\star/L_V)_{0}}} 
	 \,\approx\, 10\, S_N(t_0) \,,
\end{equation}
where $\sn(t_0) \approx 0.5$ (Harris 1991) is the Galactic \sn\ today.
This is simply meant to illustrate that after the formation of the 
Galactic GCs and before the gas settled down to form stars in the disk, 
the relative amount of stellar mass in GCs was greater.   
In this sense, \sn\ (i.e., $\sn^\prime$) was higher
at early times and has decreased because of star formation.

Of course, even if very few stars form in the cluster center after 
collapse, star formation will continue in other cluster galaxies
as in the Milky Way; given enough time, the central galaxy will
accrete enough stars (and some GCs too) to decrease 
its \sn\ (without affecting \egc).  McLaughlin \etal\ (1994) called
this proces \sn~``dilution'' and suggested that they saw evidence
for it in a possible correlation between \sn\ and BM type.
Improved measurements and larger samples (Ostrov \etal\ 1998; BTM)
fail to support this trend, but this may be because BM type
is not a robust measure of  evolutionary state.  In addition, kinematical
studies of the galaxies within cluster centers indicate that the current
merger rate extrapolated over a cluster lifetime will increase the
BCG luminosity by only $\sim 1L^*$ 
(Merrifield \& Kent 1991; Blakeslee \& Tonry 1992),
consistent with the estimates of Merritt (1985), although 
Arag\'on-Salamanca \etal\ (1998) find a higher rate from the $K$-band
Hubble diagram for BCGs out to $z{\,\sim\,}1$, approaching the cluster
formation epoch.

The time scale for such \sn\ dilution will be shortest in highly
compact, low-dispersion clusters such as Fornax.  (Indeed, given the
recent precipitous drop in published values of \sn\ for NGC~1399, one might
conclude it was on the order of years.)  BTM puzzled over Fornax because
its supposed high-\sn\ cD did not fit in with the observed
correlations.  However, two revised estimates both give $\sn =
3.9\pm0.6$ (Ostrov \etal\ 1998; Kissler-Patig \etal\ 1999), where we
have corrected to the mean Cepheid distance $\mM = 31.40$ (Ferrarese
\etal\ 1999) found for two Fornax spirals.
Thus NGC 1399 now agrees well with the relation given by Eq.\,(\ref{eq:snsig}),
for instance.  However, the predominance of the cD halo in this galaxy
and the correlation of
\sn\ with the extendedness of the light distribution (BTM) makes it
reasonable to hypothesize that NGC 1399 originally had a higher \sn\ which
has decreased through cluster dynamical evolution.  There is evidence from
X-ray data (Jones \etal\ 1997) that nearby NGC 1404 is in the process of being
accreted; when this occurs, NGC 1399's \sn\ will drop by 10\%
(see Kissler-Patig \etal\ 1999).

Finally, returning to the present sample of galaxies, we ask if there is
any evidence for \sn\ evolution here.  The sample covers the
spectrum of morphological type.
Usually differences in cluster richness obscure any effects that might
be associated with morphological state, but here the six clusters are
fairly similar in richness.  As noted earlier, the A2151 BCG has the
highest $\sn = 12.5$ and also a spiral fraction near 50\%, indicating a
significant amount of ongoing star formation in this cluster.  The A2152
BCG and its close neighbor, which together constitute a binary BCG, have
high values of \sn\ given that the X-ray properties show A2152 to be the
poorest in the sample.  The richest cluster is A754, which unexpectedly
has the lowest $\sn = 6.8$, but its morphology and the BCG's
high luminosity suggest that the cluster is well evolved.  Similarly,
the A1644 BCG's \sn\ would increase to $\sim10$ if its luminosity were
about the mean for this sample.  Thus, we may be beginning to see the
effects of central galaxy luminosity evolution causing a decrease in
\sn, although much work is needed before this can be confirmed.

\subsection{The GCLF and Cluster Distances}
\label{ssec:discgclf}

BTM noted that their results were consistent with the GCLFs in their sample
clusters being scaled versions of the M87 GCLF translated in distance; this
supported the idea of a universal GCLF for cluster gE's.
In contrast, the recent work of Ferrarese \etal\ (1999) warns against using
the GCLF as a distance indicator because the $V$-band GCLF is intrinsically
brighter by $0.5\pm0.1$ mag in the Fornax poor cluster than in Virgo, based
on their Cepheid distances and published GCLF measurements. The situation
for the $B$-band is even worse: the GCLF is brighter by $0.64\pm0.25$ mag in
Fornax, and by nearly 1.2~mag in the Leo group, than in Virgo.  Blakeslee \&
Tonry (1996) earlier noted this trend and proposed that it was due to
environmental effects; as further evidence, they pointed to the then
preliminary HST results indicating that the GCLF in the richer Coma cluster
was significantly {\it fainter} than in Virgo.  In fact, the results of 
Baum \etal\ (1997) do indicate that the GCLF is fainter by $0.4\pm0.2$~mag
in Coma than in Virgo when compared to the relative distances from most
other methods (e.g., van den Bergh 1992; Jerjen \& Tammann 1993;
D'Onofrio \etal\ 1997; Kelson \etal\ 1999). 
As all of the measurements are for ellipticals, the problem cannot be
solved by invoking different GCLFs for different Hubble types.

Ferrarese \etal\ (1999) note that the Coma HST data do not adequately
sample the GCLF turnover and thus conclude the result is unreliable.
Moreover, the BTM results are more consistent with the GCLF being the
same in Coma as in Virgo.  The difference of $\sim0.1$~mag noted in
\S\ref{ssec:resgclf} between the mean \siglf\ of the present sample 
and that of M87 is in the wrong sense---the GCLF would have to be 
{\it brighter} by $\sim0.3$ mag in these dense clusters than in Virgo 
and the BTM clusters (although still fainter than in Fornax).  Thus,
while ellipticals in poor groups appear to have intrinsically brighter
GCLFs than those in clusters, there is no strong evidence for further
variation in the GCLF among clusters over a very large range in richness.

If we assume that the GCLF is universal for BCGs, we can use a
restricted form of the GCLF distance method for the sample galaxies.
Instead of fixing \mor\ from the CMB velocity and constraining \siglf,
we can fix \siglf\ to some value and vary \mor\ until the point source
and fluctuation measurements give the same answer for the total number
of GCs; thus we obtain an estimate of the relative cluster distances.
Of course, this approach is subject to all the same potential problems
from reaching unequal depths along possibly non-Gaussian GCLFs.  That
said, we note that five of the BCGs have \siglf\ consistent with
the median of 1.53 mag, while the A2152 BCG has
$\siglf=1.36$ mag.  This assumes that distance goes as redshift
$d\sim z$ (from Barmby \& Huchra 1998) so that A2152 is 30~Mpc behind
the other two Hercules supercluster members A2147 and A2151. 
Requiring the A2152 \siglf\ to be $\sim0.15$ mag larger to match the
others gives a distance $23$\% larger than implied by its redshift, 
so that A2152 is $\sim$60 Mpc in the background and falling towards the 
supercluster with a peculiar velocity near 2500~\kms.  Given the 
uncertainties and the sensitivity of \mor\ on changes in \siglf, this
``Hercules infall'' is clearly not a significant result, and we continue
to prefer the assumption that redshift reflects distance.

On the positive side, the GCLF analysis does indicate that A2152 is not
likely to be at the same distance as the other two Hercules clusters and
falling away from us at high speed through the ``plane'' of the supercluster.
If we do assume that it is at the same distance, then its \siglf\ drops to
$1.24\pm0.10$ mag, which is significantly less than that of any of the
other clusters, as well as less than for any of the 23 BTM galaxies.  
(Some galaxies, including M31 [Secker 1992], do have GCLFs this narrow
or narrower, but none among this large sample of similar galaxies.)
Thus, this restricted form of the GCLF method, requiring
both the mean and the dispersion to be universal for BCGs, implies that
A2152 is significantly behind the other two supercluster members,
consistent with the mean redshifts from Barmby \& Huchra (1998) and in
contrast to previous works that found roughly equal mean redshifts for
all three clusters (e.g., Zabludoff \etal\ 1993).  It would be worth
trying to disentangle this triple cluster system with more robust methods
such as the fundamental plane or Tully-Fisher.

\section{Summary and Conclusions}

Deep LRIS $R$-band imaging has been presented for the centers of six
rich Abell clusters, all with velocity dispersions 
$\sigma_{\rm cl} > 700$ \kms.  We have used both the counts of
point sources as a function of radius and the residual fluctuations in
surface brightness after the removal of these point sources to study
the GC populations around the central cluster galaxies.  
New features necessary for the analysis of the present data set include
allowances for a spatially varying \pf\ and
small corrections to the background surface densities for gravitational
lensing effects.  The point source counts give somewhat higher values
for the extrapolated total numbers of GCs, if the width of the GCLF is
$\siglf\sim1.4$; we have offered several reasons why the counts might be
biased high.  Alternatively, if there is no such bias, we have measured
a median $\siglf = 1.53$~mag.
This is about 0.1 mag broader than that found for the BTM sample
of galaxies in generally lower-mass clusters.  A restricted application
of the GCLF distance method indicates that A2152 is likely more distant
than the rest of the Hercules supercluster.

The correlations found by BTM for \sn\ with velocity dispersion and
X-ray luminosity continue to hold up for these rich clusters.
We have provided empirical scaling relations for \sn\ in the combined data set.
The derived values for the number of GCs per unit mass \egc\ are similar to
those found for the BTM sample; the implied ``universal'' GC formation
rate is $\egc\sim0.5$--1~GC per 10$^{9}\,M_\odot$.
Thus, the variation seen in \sn\ with
cluster mass seems to result from ``missing light'' rather than anything
unusual in the number of GCs.  The GC color distributions found by other
authors in central cluster galaxies lend further support to this view.

Guided by recent numerical simulations of BCG formation, we have argued
that the observations are explained by a scenario in which the GCs formed
at early times and fell into the cluster center along with their associated
galaxies which merged to form the BCG.  In this process of cluster collapse
and BCG formation, the gas may have become heated and lost to the
intracluster environment (BTM, Harris \etal\ 1998) so that star formation
ceased early in the cluster center.  Although the simulations do not yet
predict the star formation rates, they indicate that the BCG formation process
is more rapid in richer clusters, consistent with the above scenario for
producing higher-\sn, higher-$M/L$ central galaxies in these clusters.

Over the subsequent lifetime of the cluster, a high-\sn\ central galaxy
will slowly grow in luminosity and evolve towards lower \sn.  
In particular, continuing star formation in the general galaxy population
of a young, spiral-rich cluster may allow the central galaxy to grow 
more luminous without the addition of a significant number of new GCs.  
We have suggested that the available data are beginning to show the first
evidence for such evolution: the most spiral-rich, irregular cluster in our
sample has the BCG with highest \sn, and the most centrally concentrated,
apparently well-evolved clusters have BCGs with lower values of \sn\ and
higher luminosities.

\acknowledgments

I wish to thank Judy Cohen, Bev Oke, and the rest of the team
responsible for producing the Low Resolution Imaging Spectrograph.  I
thank Pat C\^ot\'e for reading and commenting on the manuscript, Mark
Metzger for his help with DoPhot, John Tonry for the use of much software,
and Judy Cohen for helpful comments throughout this work.
G.~Kochanski made some enlightening comments at the AAS Meeting in
Austin, and Terry Stickel provided excellent assistance at the telescope.
I also thank the anonymous referee for helpful comments and corrections.
This research has made use of the NASA/IPAC Extragalactic Database (NED),
which is operated by the Jet Propulsion Laboratory at Caltech, under
contract with the National Aeronautics and Space Administration. 
I gratefully acknowledge the financial support of
a Sherman M.~Fairchild Fellowship.

\setcounter{figure}{0}  \vskip 2.5 truein 
\noindent{\bf Captions for GIF images of program fields:}\medskip
\figcaption{\small Center of the Keck LRIS $R$-band image
of Abell~754. North is up; East is to the left. \label{img:a754}} 
\figcaption{\small Center of the Keck LRIS $R$-band image 
of Abell~1644. North is up; East is to the left.\label{img:a1644}}
\figcaption{\small Center of the Keck LRIS $R$-band image 
of Abell~2124. North is up; East is to the left.\label{img:a2124}}
\figcaption{\small Center of the Keck LRIS $R$-band image 
of Abell~2147. North is up; East is to the left.\label{img:a2147}}
\figcaption{\small Center of the Keck LRIS $R$-band image 
of Abell~2151. North is up; East is to the left.\label{img:a2151}}
\figcaption{\small Center of the Keck LRIS $R$-band image 
of Abell~2152. North is up; East is to the left.\label{img:a2152}}

\clearpage

\begin{deluxetable}{ccccccc}\tablenum{3}\small
\centering\tablewidth{0pt}\tabcolsep=0.42cm
\newdimen\digitwidth \setbox0=\hbox{\rm0} \digitwidth=\wd0 \catcode`?=\active
\def?{\kern\digitwidth}
\tablecaption{Power-Law Fits to GC Radial Number Density Distributions\label{tab:nrad}}
\tablehead{
\colhead{Field} & $m_b,\, m_c$ & $ R_{\rm max}$ & 
\colhead{$\log A$\rlap{$_0$}} & \colhead{$\pm$} & \colhead{$\;\;\alpha$} & \colhead{$\pm$} 
} \startdata
  A754 & $23.0,\,25.6$ & 140\arcsec\ & 3.29 & 0.32  & $-$1.02 & 0.22 \\
 a1644 & $23.0,\,26.0$ & 160\arcsec\ & 4.62 & 0.13  & $-$1.56 & 0.09 \\
 a2124 & $23.6,\,26.6$ & 120\arcsec\ & 5.22 & 0.20  & $-$1.96 & 0.16 \\
 a2147 & $22.5,\,25.6$ & 215\arcsec\ & 4.29 & 0.13  & $-$1.40 & 0.10 \\
 a2151 & $22.5,\,25.8$ & 205\arcsec\ & 4.48 & 0.12  & $-$1.35 & 0.09 \\
 a2152 & $22.5,\,26.0$ & 175\arcsec\ & 4.62 & 0.54  & $-$1.66 & 0.34 \\
\enddata\vspace{-0.6cm}
\tablecomments{Table lists for each cluster: the bright and faint limiting $R$-band
mag\-nitudes $m_b,\,m_c$ of the background-corrected point source counts used for the fit;
radius corresponding to 105~$h^{-1}\,$kpc out to which the fits were done;
base-ten logarithm of the best-fit scale factor $A_{0}$ 
(which has units of number/arcmin$^2$/arcsec$^{\alpha}$) 
and its uncertainty; and best-fit power-law exponent $\alpha$ and its uncertainty.
}
\end{deluxetable}

\begin{deluxetable}{rcrrcrrcrrrr}\tablenum{4}
\centering \tablewidth{0pt} \tabcolsep=0.3cm
\tablecaption{Point Source Counts and Variance Measurements\label{tab:data}}
\small 
\newdimen\digitwidth \setbox0=\hbox{\rm0} \digitwidth=\wd0 \catcode`?=\active
\def?{\kern\digitwidth}
\tablehead{
\colhead{Galaxy.reg} & \colhead{?$m_b$} & \colhead{$N_{\rm ps}$} & \colhead{$\pm$} &
\colhead{$f_\ell$} & \colhead{$N_{\rm GC}$} & \colhead{$\pm$} & \colhead{$m_c$} & \colhead{$P_0$} &
\colhead{$\pm$} & \colhead{$P_{\rm GC}$} & \colhead{$\pm$} }
\startdata
A754-1.c1   & 23.0 &  175.7 & 106.4 & 0.77 & 146.4 & 106.5 & 25.4 & 1152 & 179 & 1026 & 180 \nl
A754-1.c2   & 23.0 &  110.1 &  17.9 & 0.85 &  77.7 &  18.5 & 25.4 &  637 &  46 &  503 &  49 \nl
A754-1.c3   & 23.0 &   92.6 &   7.9 & 0.93 &  48.1 &   9.8 & 25.6 &  353 &  24 &  236 &  28 \nl
A754-1.c4   & 23.0 &   68.9 &   3.5 & 0.98 &  22.3 &   6.9 & 25.6 &  183 &  14 &   62 &  20 \nl
A1644-1.c1  & 23.0 &  383.2 &  69.1 & 0.80 & 357.1 &  69.1 & 25.6 & 2182 & 142 & 1999 & 144 \nl
A1644-1.c2  & 23.0 &  267.3 &  32.4 & 0.86 & 239.1 &  32.5 & 25.6 & 1279 &  83 & 1104 &  85 \nl
A1644-1.c3  & 23.0 &  194.1 &  11.8 & 0.94 & 139.1 &  12.2 & 26.0 &  418 &  28 &  300 &  30 \nl
A1644-1.c4  & 23.0 &  101.7 &   4.1 & 0.98 &  44.5 &   5.0 & 26.0 &  227 &  15 &  107 &  19 \nl
A2124-1.c1  & 23.6 &  641.6 & 125.2 & 0.73 & 595.8 & 125.3 & 26.1 & 2675 & 241 & 2446 & 243 \nl
A2124-1.c2  & 23.6 &  390.7 &  37.1 & 0.85 & 323.0 &  37.4 & 26.3 & 1287 & 102 & 1087 & 104 \nl
A2124-1.c3  & 23.6 &  235.9 &  13.4 & 0.94 & 133.4 &  13.8 & 26.6 &  431 &  29 &  276 &  32 \nl
A2124-1.c4  & 23.6 &  124.1 &   4.4 & 0.98 &  17.4 &   5.4 & 26.6 &  196 &  13 &   36 &  19 \nl
A2147-1.c1  & 22.5 &  322.2 &  70.1 & 0.88 & 287.9 &  70.1 & 25.3 & 1195 & 121 & 1086 & 122 \nl
A2147-1.c2  & 22.5 &  261.2 &  25.4 & 0.91 & 222.2 &  25.5 & 25.4 &  555 &  40 &  467 &  42 \nl
A2147-1.c3  & 22.5 &  131.9 &   9.4 & 0.95 &  91.1 &   9.7 & 25.4 &  326 &  21 &  240 &  26 \nl
A2147-1.c4  & 22.5 &   90.3 &   3.9 & 0.98 &  38.8 &   5.0 & 25.6 &  143 &   9 &   72 &  15 \nl
A2151-1.c1  & 22.5 &  211.0 &  62.6 & 0.92 & 193.9 &  62.6 & 25.0 & 1241 &  88 & 1143 &  88 \nl
A2151-1.c2  & 22.5 &  285.2 &  31.0 & 0.93 & 258.3 &  31.2 & 25.4 &  553 &  39 &  490 &  39 \nl
A2151-1.c3  & 22.5 &  252.9 &  13.3 & 0.96 & 207.4 &  14.2 & 25.8 &  186 &  12 &  145 &  13 \nl
A2151-1.c4  & 22.5 &  130.7 &   4.6 & 0.99 &  84.2 &   6.7 & 25.8 &   92 &   6 &   51 &   7 \nl
A2152-1.c1  & 22.5 &  112.3 &  32.9 & 0.90 &  86.0 &  33.0 & 25.5 &  192 &  16 &  165 &  16 \nl
A2152-1.c2  & 22.5 &  153.7 &  21.2 & 0.92 & 101.7 &  21.6 & 26.0 &   55 &   5 &   40 &   5 \nl
A2152-1.c3  & 22.5 &  130.6 &   9.7 & 0.96 &  76.5 &  10.5 & 26.0 &   38 &   3 &   23 &   4 \nl
A2152-1.c4  & 22.5 &   86.4 &   3.8 & 0.99 &  31.0 &   5.7 & 26.0 &   23 &   2 &    8 &   2 \nl
A754-2.c1   & 23.0 &  296.3 & 141.8 & 0.98 & 258.9 & 141.9 & 25.4 &  784 & 180 &  632 & 181 \nl
A754-2.c2   & 23.0 &   89.2 &  15.9 & 0.98 &  51.9 &  16.6 & 25.4 &  389 &  28 &  239 &  34 \nl
A754-2.c3   & 23.0 &   66.6 &   6.6 & 0.98 &  19.8 &   8.9 & 25.6 &  194 &  26 &   72 &  29 \nl
A2151-2.c1  & 22.5 &  175.7 &  40.2 & 0.99 & 157.3 &  40.3 & 25.0 &  725 &  65 &  623 &  65 \nl
A2151-2.c2  & 22.5 &  153.8 &  20.6 & 0.99 & 125.3 &  20.9 & 25.4 &  262 &  18 &  198 &  19 \nl
A2151-2.c3  & 22.5 &  141.3 &   9.7 & 0.99 &  94.6 &  10.8 & 25.8 &  110 &   7 &   69 &   8 \nl
A2152-2.c1  & 22.5 &  169.8 &  44.7 & 0.97 & 141.5 &  44.8 & 25.5 &  207 &  17 &  180 &  17 \nl
A2152-2.c2  & 22.5 &  224.8 &  26.8 & 0.97 & 170.3 &  27.1 & 26.0 &   63 &   5 &   48 &   5 \nl
A2152-2.c3  & 22.5 &  127.9 &  10.2 & 0.97 &  73.4 &  11.0 & 26.0 &   35 &   2 &   20 &   3 \nl
\enddata\vspace{-0.6cm}
\tablecomments{For each annular region of each galaxy, the table lists:
bright cutoff magnitude $m_b$ of the point source counts;
incompleteness-corrected number of point sources $N_{\rm ps}$ (arcmin$^{-2}$) 
fainter than $m_b$ but brighter than $m_c$; the lensing factor $f_\ell$ that
was applied to the background estimate; total number of GCs $N_{\rm GC}$ (arcmin$^{-2}$)
in the $m_b$-$m_c$ magnitude range; faint cutoff $m_c$;
fluctuation power $P_0$ from objects fainter than $m_c$,
in units of $10^4\,$($e^-$/pixel)$^2$;
the power $P_{\rm GC}$ due to GCs fainter than $m_c$,
also in $10^4\,$($e^-$/pixel)$^2$. 
}
\end{deluxetable}

\begin{deluxetable}{rccccccc}\tablenum{5}
\centering 
\tablewidth{0pt}\tabcolsep=0.25cm
\tablecaption{Results for $\sigma_{\sc lf}$ and $S_N$\label{tab:Sn}}
\newdimen\digitwidth \setbox0=\hbox{\rm0} \digitwidth=\wd0 \catcode`?=\active
\def?{\kern\digitwidth}
\tablehead{ 
\tablevspace{-6pt}
\colhead{} & \colhead{} & \colhead{} & \colhead{?$_{(\sigma_{\sc lf}{=\,}1.4)}$} &
\colhead{?$_{(\sigma_{\sc lf}{=\,}1.5)}$} & \colhead{?$_{(\sigma_{\sc lf}{=\,}1.6)}$} &
\colhead{} & \colhead{} \nl
\colhead{Galaxy} & \colhead{?$M_V^{40}$} & \colhead{$\sigma_{\sc lf}$~~~$\pm$} &
\colhead{?$S_{N}^{40}\;\;\;^{+}_{-}$} & \colhead{?$S_{N}^{40}\;\;\;^{+}_{-}$} & 
\colhead{?$S_{N}^{40}\;\;\;^{+}_{-}$}
& \colhead{?$M_V^{65}$} & \colhead{?$S_{N}^{65}\;^{+}_{-}$}
}\startdata
 A754-1 &$-$22.87 & 1.54~~0.10 & ?6.8~~0.5 & ?5.8~~0.4 & ?4.9~~0.4 &$-$23.31 &?$6.8\;^{2.3}_{1.7}$
\nl\tablevspace{5pt}
A1644-1 &$-$23.04 & 1.60~~0.09 & ?6.5~~0.3 & ?5.8~~0.3 & ?5.2~~0.2 &$-$23.36 &?$7.0\;^{2.0}_{1.6}$
\nl\tablevspace{5pt}
A2124-1 &$-$22.71 & 1.64~~0.14 & ?9.3~~0.5 & ?8.2~~0.4 & ?7.3~~0.4 &$-$23.00 &?$9.9\;^{2.9}_{2.3}$
\nl\tablevspace{5pt}
A2147-1 &$-$22.30 & 1.53~~0.09 & ?9.6~~0.5 & ?8.6~~0.4 & ?7.7~~0.4 &$-$22.67 &?$9.8\;^{2.8}_{2.2}$
\nl\tablevspace{5pt}
A2151-1 &$-$22.42 & 1.50~~0.09 & 12.5~~0.5 & 11.2~~0.4 & 10.1~~0.4 &$-$22.85 &$12.5\;^{3.4}_{2.7}$
\nl\tablevspace{5pt}
A2152-1 &$-$21.92 & 1.36~~0.10 & ?7.8~~0.5 & ?6.9~~0.5 & ?6.1~~0.4 &$-$22.19 &?$9.1\;^{2.7}_{2.1}$
\nl\tablevspace{5pt}
 A754-2 &$-$21.36 & 1.72~~0.22 & 12.3~~1.9 & 10.6~~1.6 & ?9.2~~1.4 & ?\dots  & ?\dots \nl\tablevspace{3pt}
A2151-2 &$-$21.62 & 1.55~~0.11 & 10.3~~0.6 & ?9.4~~0.5 & ?8.6~~0.5 & ?\dots  & ?\dots \nl\tablevspace{3pt}
A2152-2 &$-$21.43 & 1.47~~0.11 & 12.7~~0.8 & 11.5~~0.7 & 10.4~~0.7 & ?\dots  & ?\dots \nl\tablevspace{3pt}
\enddata\vspace{-0.6cm}
\tablecomments{Columns list for each galaxy: absolute $V$-band magnitude $M_V^{40}$
projected within 40 kpc ($h{=}0.8$); best-fit Gaussian width
$\sigma_{\sc lf}$ of the GC luminosity function (mag); metric specific
frequencies $S^{40}_{N}$ within projected radii of 40 kpc and their
internal errors for $\sigma_{\sc lf}$ values of 1.40, 1.50, and 1.60
mag; absolute $V$-band magnitude $M_V^{65}$ projected within 65 kpc; and
metric specific frequencies $S^{65}_{N}$ and total uncertainties
within projected radii of 65 kpc ($\sigma_{\sc lf} = 1.45\pm0.05$).  }
\end{deluxetable}

\begin{deluxetable}{rcccc}\tablenum{6}
\centering 
\tablewidth{0pt}
\tablecaption{Estimates of GCs per Mass\label{tab:egc}}
\tabcolsep=0.3cm
\newdimen\digitwidth \setbox0=\hbox{\rm0} \digitwidth=\wd0 \catcode`?=\active
\def?{\kern\digitwidth}
\tablehead{ 
\colhead{} & \multicolumn{2}{c}{singular} & \multicolumn{2}{c}{flat core} \nl
\colhead{Cluster} & \colhead{$\eta_{\sc gc}^{40}$} &  \colhead{$\eta_{\sc gc}^{65}$}
& \colhead{$\eta_{\sc gc}^{40}$} &  \colhead{$\eta_{\sc gc}^{65}$} 
}\startdata
  A754  & 0.44 & 0.42 & 0.99 & 0.60 \nl
 A1644  & 0.51 & 0.45 & 1.15 & 0.66 \nl
 A2124  & 0.57 & 0.48 & 1.23 & 0.68 \nl
 A2147  & 0.43 & 0.38 & 0.95 & 0.54 \nl
 A2151  & 0.85 & 0.78 & 1.71 & 1.01 \nl
 A2152  & 0.33 & 0.31 & 0.66 & 0.39 \nl
\enddata\vspace{-0.6cm}
\tablecomments{The second and third columns give the number of
GCs per unit $10^9\,M_\odot$ within 40 and 65 kpc, respectively, 
as estimated from a singular isothermal cluster model.
The last two columns give the same quantities but estimated
from a flat core model as in BTM.
}
\end{deluxetable}

\end{document}